\documentclass[aps,prd,twocolumn]{revtex4-2}

\usepackage{bm}
\usepackage{amsmath}
\usepackage{graphicx}
\usepackage{subcaption}

\usepackage{changes}
\usepackage{latexsym}
\usepackage{amsfonts}
\usepackage{mathrsfs}
\usepackage{CJKutf8}

\usepackage{float}
\usepackage{color}
\usepackage{comment}
\usepackage{hyperref}

\hypersetup{
	colorlinks=true,
	linkcolor=blue,
	citecolor=blue,
}

\newcommand{\dm}{\mathrm{d}}

\newcommand{\physrep}{Physics Reports} 
\begin{document}

\begin{CJK*}{UTF8}{gbsn} 
	
\title{Detectability of dark matter density distribution via gravitational waves 
from\\
binary black holes in the Galactic center} 
	
\author{Zhijin Li$^{1,4,5}$}
\author{Xiao Guo$^{1,6}$}
\email[corresponding author:~]{guoxiao17@mails.ucas.ac.cn}
\author{Zhoujian Cao$^{1,3}$}
\email[corresponding author:~]{zjcao@amt.ac.cn}
\author{Yun-Long Zhang$^{2,1}$}
\email[corresponding author:~]{zhangyunlong@nao.cas.cn}

\affiliation{$^{1}$School of Fundamental Physics and Mathematical Sciences, Hangzhou Institute for Advanced Study, University of Chinese Academy of Sciences, 
Hangzhou 310024, China}

\affiliation{$^{2}$National Astronomical Observatories, Chinese Academy of Sciences,
Beijing 100101, China}

\affiliation{$^{3}$Institute of Applied Mathematics, Academy of Mathematics and Systems Science, Chinese Academy of Sciences, 
Beijing 100190, China}

\affiliation{$^{4}$ International Centre for Theoretical Physics Asia-Pacific (ICTP-AP),
University of Chinese Academy of Sciences, Beijing 100049, China}

\affiliation{$^{5}$CAS Key Laboratory of Theoretical Physics, Institute of Theoretical Physics, Chinese Academy of Sciences, Beijing 100190, China.}

\affiliation{$^{6}$Institute for Gravitational Wave Astronomy, Henan Academy of Sciences, Zhengzhou 450046, Henan, China}

\begin{abstract}
The fundamental nature of dark matter (DM) remains unknown, with significant uncertainties in its density profile. DM environments surrounding massive binary black holes (BBHs) modify their orbital dynamics, thereby altering gravitational wave (GW) emissions. For BBH systems at the Galactic Center, dynamical friction induced by DM spikes could produce detectable deviations in GW spectra, potentially observable by future space-based detectors.
To address the uncertainties in the Galactic Center's DM profile, we systematically examine two scenarios: the generalized Navarro-Frenk-White (gNFW) profile and its post-spike modification. We investigate the evolutionary effects of DM dynamical friction and accretion on the eccentricity and semi-latus rectum of secondary black holes (BHs) in elliptical orbits. By constructing orbital models with varying initial eccentricities across the mass-semi-latus rectum parameter space and utilizing 30 years of simulated pulsar timing array data from the Square Kilometer Array (SKA), we identify detectable parameter regimes of DM effects and employ these GW observational signatures to constrain different DM density profiles.
Our analysis reveals that among gNFW profiles ($\gamma=2,1.5,1,0.5$), only $\gamma=2$ produces significant detectable signatures. The formation of DM spikes further enhances these observable waveform deviations for all gNFW slopes.
\end{abstract}


\maketitle
	
\end{CJK*}

\tableofcontents

\allowdisplaybreaks

\section{Introduction}

A wealth of observational evidence provides robust support for the existence of dark matter (DM)~\cite{Rubin:1970zza,Clowe:2006eq,Planck:2015fie}. Nevertheless, its fundamental nature remains one of the most pressing and profound mysteries in modern physics, while its density profile continues to be a subject of intense scientific interest.
Through the cosmological $N$-body numerical simulations,
Navarro, Frenk and White (NFW) obtained a universal density profile for DM halos, called NFW profile~\cite{Navarro:1995iw,Navarro:1996gj}. 
The NFW profile exhibits a density scaling of $\rho(r)\sim r^{-1}$ at small scales, while subsequent studies suggested a steeper slope of $\rho(r)\sim r^{-1.5}$ ~\cite{Moore:1999gc,Jing:1999ir}, leading to cuspy distributions in the central regions. However, there is no conclusive evidence for such steep inner density slopes, the simulations for the NFW profile and the Einasto profile do not match well on small scales~\cite{Wang:2019ftp}. Instead, observations favor the existence of a ``core" where the DM distribution becomes relatively flat, the so-called core-cusp problem~\cite{Flores:1994gz, Moore:1994yx,2018PhR...730....1T}.

It has been demonstrated that if a supermassive black hole (SMBH) (with mass regime $10^6$--$10^9\,M_{\odot}$) is embedded in a DM halo~\citep{Gondolo:1999ef}, its adiabatic compression of the surrounding DM inevitably produces a high-density, concentrated spike~\citep{Gondolo:1999ef}. The resulting density profile follows $\rho(r) \sim r^{-\gamma_{\rm sp}}$, with a significantly steeper slope in the range $2.25 \le \gamma_{\rm sp} \le 2.5$. Besides, the dynamical processes such as major merger events of seed host galaxies and gravitational scattering by stars may lead to the destruction or reduction of DM spikes~\cite{Ullio:2001fb,Merritt:2002vj,merritt2004evolution,DeLuca:2023laa}. 
The Galactic Center (GC), our closest galactic nucleus, hosts a SMBH (Sgr~A*) ~\cite{balick1974intense,Schodel:2002py,Ghez:2003rt,Genzel:2010zy}. This unique astrophysical laboratory provides unprecedented opportunities to study DM properties/distributions and test general relativity. The DM distribution in the inner Galactic region remains poorly understood due to the lack of direct observational data and the limited resolution of numerical simulations (below $\sim 1\,\mathrm{kpc}$). Recent studies suggest that Sgr A$^*$ has not experienced major mergers in the past $10\,\mathrm{Gyr}$ ~\cite{micic2007supermassive,wang2024evidence}, making the existence of a DM spike around Sgr A$^*$ uncertain. 
Although the Milky Way's (MW) DM distribution has been investigated through rotation curve analyses ~\cite{hooper2011dark,ackermann2015updated,di2021characteristics,cholis2022return}, constraining its inner profile (within $\sim\mathrm{kpc}$ scales) remains challenging ~\cite{sofue2013rotation}. The distribution of DM in the inner Galactic region remains an open question, as the environment near the central BH is highly complex and may involve interactions with baryonic matter. Additionally, the nature of DM itself introduces significant uncertainties in its density profile.

On the other hand, the intermediate-mass black holes (IMBHs) may exist around Sgr A* within certain parameter ranges ~\cite{yu2003ejection,hansen2003need,Genzel:2010zy,girma2019astrometric,naoz2019hidden,abuter2020detection,straub2023intermediate,will2023constraining}. Such an IMBH would form a binary black hole (BBH) system with Sgr A*, with orbital periods ranging from months to years. The resulting gravitational wave (GW) emission falls within the nHz-$\mu$Hz frequency range, detectable by pulsar timing arrays (PTAs) ~\cite{maggiore2008gravitational,sesana2009gravitational,sesana2010gravitational,lee2011gravitational,sesana2013gravitational,zhu2014all,schutz2016constraints,taylor2019supermassive,agazie2023nanograv,antoniadis2024second}. 
The GW signals from this system may exhibit detectable differences due to DM environmental effects compared to DM-free scenarios, potentially serving as a probe for mapping the DM distribution in the GC. 
Since 2015, LIGO's GW detections have inaugurated the era of GW astronomy ~\cite{LIGOScientific:2016aoc,LIGOScientific:2017vwq,LIGOScientific:2017ync,LIGOScientific:2018mvr}. Recent observations by the NANOGrav ~\cite{NANOGrav:2023gor,NANOGrav:2023hvm}, Parkes PTA ~\cite{Reardon:2023gzh}, Chinese PTA ~\cite{Xu:2023wog}, and European PTA ~\cite{EPTA:2023fyk} collaborations have reported evidence (2-4\,$\sigma$ confidence level) for a stochastic gravitational wave background (SGWB) in the nanohertz frequency range. These findings present both new opportunities and challenges ~\cite{Bertone:2019irm} for investigating DM distribution in the GC region. The future Square Kilometer Array (SKA) is projected to detect new stable millisecond pulsars (MSPs) and form a high-sensitivity PTA~\cite{lazio2013square,wang2017pulsar}, designated as the SKA-PTA. In the near future, GW  observations could potentially reveal the DM distribution in the GC.

The presence of DM subjects compact objects moving through it to dynamical friction~\cite{Chandrasekhar:1943ys,Ostriker:1998fa}. Eda et al.\cite{Eda:2013gg,Barausse:2014tra,Eda:2014kra} pioneered the investigation of how dynamical friction from DM mini-spikes affects intermediate-mass-ratio inspirals (IMRIs) - GW sources potentially detectable by LISA in the future. Macedo et al.\cite{Macedo:2013qea} examined a compact object traversing a DM region, where the quasi-adiabatic inspiral is primarily dominated by dynamical friction and accretion rather than GW radiation. An IMRI system consisting of an intermediate-mass black hole (IMBH) and a smaller BH, when accounting for GW radiation, dynamical friction, and accretion within the minispike, can influence the GW phase and inspiral period \cite{Yue:2017iwc}. The dynamical friction from a DM minispike within an IMRI system will increase the eccentricity of the elliptical orbit~\cite{Yue:2019ozq}.  A halo feedback mechanism is proposed in which a secondary object loses energy while moving through a DM spike, depositing this energy into the surrounding DM. This process reduces the DM density, consequently weakening the dynamical friction effect \cite{Kavanagh:2020cfn}. Similarly, the accretion of DM by the secondary BH within the spike can lead to significant dephasing and disrupt the spike structure, resulting in an accretion feedback effect~\cite{Karydas:2024fcn}.

The presence of a DM minispike in an IMRI system introduces additional orbital precession, leading to a detectable phase shift in the GW waveform \cite{Dai:2021olt}. Taking into account the velocity distribution of DM particles relative to the small BH, the dynamical friction induced by the DM spike would circularize the elliptical orbit~\cite{Becker:2021ivq}. In extreme-mass-ratio inspirals (EMRIs), dynamical friction can significantly alter gravitational waveforms in the frequency domain~\cite{Li:2021pxf,2025arXiv250602937F,2024JCAP...01..035R}, while a DM halo may reduce precession rates or even reverse their direction\cite{Dai:2023cft}. Reference ~\cite{Speeney:2022ryg} investigated the influence of DM distributions under relativistic conditions on BBH systems, ~\cite{Becker:2022wlo,Cole:2022yzw} examined the effects of baryonic matter accretion disks, ~\cite{Shadykul:2024ehz,2025arXiv250504697M} analyzed the case of slowly rotating Kerr BHs, while ~\cite{daniel2025forecasted} studied the impact of DM spikes surrounding SMBHs on inspiral dynamics.

In this paper, we investigate the dynamical evolution of secondary BHs orbiting within the DM halo surrounding the SMBH at the GC. Their orbital dynamics are influenced by both dynamical friction and DM accretion effects. Focusing on the GC's DM density profile, we address current uncertainties in its inner slope and the potential existence of a DM spike by comparing two scenarios: (1) a generalized NFW profile without spike ($\gamma=0.5-2.0$), and (2) a spike-modified profile accounting for DM enhancement. 
We systematically analyze how these DM distributions affect the orbital evolution (eccentricity and semi-latus rectum) and corresponding GW signatures. By exploring the parameter space of secondary BH masses and orbital parameters, we quantify DM-induced deviations in GW waveforms compared to DM-free scenarios. These findings could enable measurement of the DM density slope, providing constraints on DM models and insights into the nature of DM. All data produced by this paper is availiable on \footnote{\url{https://github.com/lizhijin624/Data-on-detecting-dark-matter-density-with-gravitational-waves.git}}.

The paper is organized as follows. In Section \ref{sec:DM}, we introduce the generalized NFW profile for the MW's DM distribution and subsequently derive its spike-like modification resulting from adiabatic compression within the gravitational influence radius of the SMBH. 
In Section \ref{sec:per}, we examine two distinct dynamical effects: DM accretion by secondary BHs and the resulting orbital element evolution induced by DM dynamical friction. In Section \ref{sec:waveform}, we explore the observable parameter space where DM effects on secondary BHs become significant and evaluate the prospects for detecting these effects in future GW observations. Finally, we draw conclusions and discussion in Section \ref{sec:concl}. 

\section{Dark Matter Density Profile}
\label{sec:DM}

In this section, we investigate the SMBH at the center of the MW galaxy, which is enveloped by a DM halo. Since the internal DM distribution remains uncertain, particularly regarding the presence or absence of a spike, we consider two distinct DM density profiles: a gNFW profile without a spike and a spike-modified profile.

\subsection {Generalized NFW profile}
\label{gNFW}

\begin{figure}
    \centering
    \includegraphics[height=0.2\textheight]{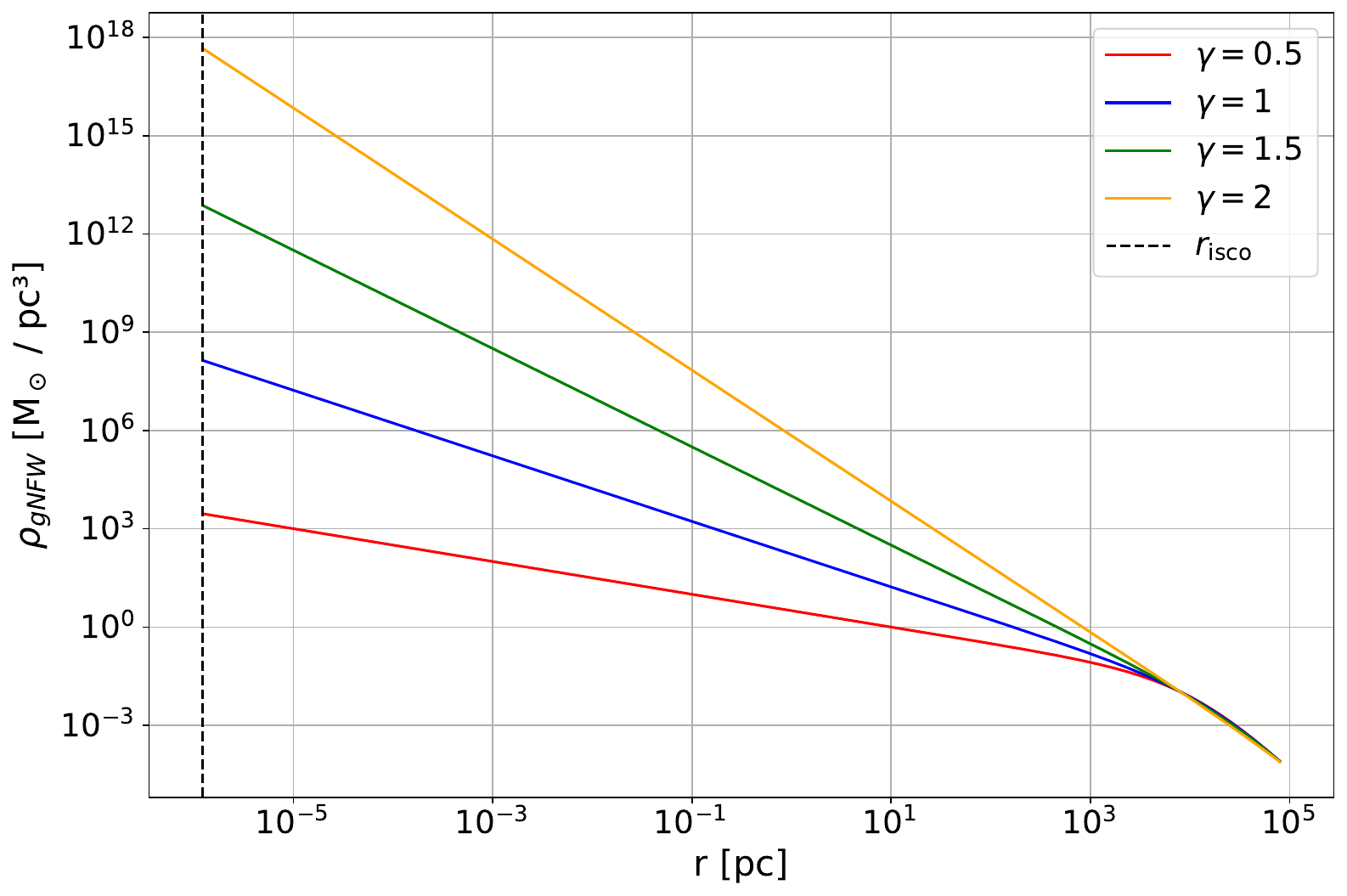}
    \caption{The DM density profiles for the gNFW halo in Eq.~\eqref{eq::gnfw_halo} are illustrated in the figure. The possible gNFW profile of DM in our MW is represented by a solid red line for $\gamma = 0.5$, a blue line for $\gamma = 1$, a yellow line for $\gamma = 1.5$, and a yellow line for $\gamma = 2$. At the center of the MW, there exists a SMBH with a mass of \(4.26 \times 10^6 \, M_{\odot}\). As one approaches the GC, the DM distribution becomes increasingly dense for different indices. However, within the innermost stable circular orbit, the DM distribution is assumed to be zero.}
    \label{fig:gNFW_profile}
\end{figure}

The DM halo density profiles are often approximated by the NFW profile~\cite{Navarro:1996gj}
\begin{equation}\label{eq:nfw_halo}
    \rho_{\rm NFW}(r)=\frac{\rho_0}{(r/r_{\rm s})(1+r/r_{\rm s})^2},
\end{equation}
where $r_{\text{s}}$  represents the typical scale radius, $\rho_{\text{0}}$ denotes the typical scale  density. One has to determine the parameters $r_{\text{s}}$ and $\rho_{\text{0}}$  that enter in the tentative DM distributions $\rho(r)$. Various methodologies exist for determining the distribution of DM within galaxies. One such approach involves extracting DM parameters from numerical simulations of galactic halos. Alternatively, the DM profile can be derived by imposing specific constraints on the galactic structure. In this study, we employ the latter method, ensuring that our model aligns with empirical data and theoretical expectations. The distance from the Sun to the center of the MW galaxy is about 8 kpc (kiloparsecs)~\cite{ghez2008measuring,GRAVITY:2021xju}, and the DM density near the Sun is ~\cite{sofue2020rotation,sofue2013rotation,cirelli2024dark}
\begin{equation}\label{eq::rho_sun}
\rho({r_{\odot}})=\rho_{\odot} = 0.4 \, \text{GeV/cm}^3 \approx 0.0106 \, M_{\odot}/ \text{kpc}^3.
\end{equation}
The total mass of DM within the MW, including that contained within the virial radius $r_{200}$ is 
~\cite{Cautun:2019eaf,cirelli2024dark}

\begin{equation}\label{M_200}
M_{DM}^{200}\approx 1.0\times 10^{12} \, M_{\odot}.
\end{equation}
For the MW,  virial radius $r_{200}$  is approximately 200 to 300 kpc. Therefore, we can consider $M_{DM}^{200}$ as the total DM mass contained within a large sphere of radius 200 kpc, centered around the center of the Galaxy. We take the DM density around the Sun and the DM contained within the virial radius of the MW (approximately 200 kpc) with a mass of $10^{12}M_{\odot}$ as constraints. In order to consider various models of DM distribution, we generalize the DM distribution to the gNFW (generalized NFW) profile~\cite{Diemand:2008in}
\begin{equation}\label{eq::gnfw_halo}
    \rho_{\rm gNFW}(r)=\frac{\rho_0}{(r/r_{\rm s})^{\gamma}(1+r/r_{\rm s})^{3-\gamma}}.
\end{equation}
Using the constraints provided by Eqs.~\eqref{eq::rho_sun} and \eqref{M_200}, we obtain the following two equations 
\begin{equation}\label{eq:gNFW_rho_sun}
\rho_{\rm gNFW}(r_{\odot}) = 0.0106  M_{\odot}/\text{kpc}^3,
\end{equation}
and

\begin{equation}\label{eq:gNFW_inte_M}
4\pi\int_{r_{\rm ISCO}}^{200 \text{kpc}} r^2\rho_{\rm gNFW}(r)  {{\dm}r}= 1.0 \times 10^{12}  M_{\odot}.
\end{equation}
Here, $r_{\text{ISCO}}$ represents the innermost stable circular orbit (ISCO) of the central SMBH, given by
$r_{\text{ISCO}} = 3R_{\text{s}} = \frac{6Gm_1}{c^2}$,
where $m_1 = 4.26 \times 10^6\, M_{\odot}$ is the mass of Sgr A$^\ast$~\cite{abuter2019geometric,schodel2002star,Schodel:2002py,Ghez:2003rt,GRAVITY:2021xju}. In this context, for the gNFW profile, we adopt a range for $\gamma$ such that $0.5 \leq \gamma\leq 2$. The DM density inside  $r_{\text{ISCO}}$ is assumed to be zero. Using Eqs.~\eqref{eq:gNFW_rho_sun} and \eqref{eq:gNFW_inte_M}, we solve for the indices
$\gamma=0.5,1,1.5,2$ and compute the corresponding values of $r_s$ and $\rho_0$, as shown in Table \ref{tab1}.
Following this, we use the data from the Table \ref{tab1} to plot the gNFW density profiles for various slopes, which are displayed in Fig. \ref{fig:gNFW_profile}.

\begin{table}[htbp]
    \centering
    \begin{tabular}{ccccc}
        \hline \hline
        \(\gamma\) & 0.5      & 1        & 1.5      & 2         \\ \hline
        \(r_s\) (pc) & 12754.12 & 19191.47 & 36818.52 & 181453.85 \\ \hline
        \(\rho_0\) (\(M_\odot/\text{pc}^3\)) & 0.0283   & 0.00887  & 0.00144  & 0.0000215 \\ \hline\hline
    \end{tabular}
    \caption{Values of the parameters \(r_s\) and \(\rho_0\) of the gNFW profile in Eq.~\eqref{eq::gnfw_halo} for different power-law indices \(\gamma\), obtained by solving Eqs.~\eqref{eq:gNFW_rho_sun} and \eqref{eq:gNFW_inte_M}.}    
    \label{tab1}
    \captionsetup{justification=raggedright, singlelinecheck=false}
    
\end{table}

\subsection{Spike profile}
\label{spike}
Gondolo and Silk~\cite{Gondolo:1999ef} pointed out that the adiabatic growth of the SMBH at the center of the DM halo will significantly increase the DM density within its gravitational influence radius $r_{\text{sp}}\sim 0.2r_h$~\cite{Merritt:2003qc}. 
Here, $r_h$ is the radius of gravitational influence of the SMBH, defined by the equation $4\pi \int_{0}^{r_{h} } \rho \left ( r \right ) r^{2}{{\dm}r}=2m_1$, where $m_1$ is  mass of Srg A*. In this study, we assume that the DM distribution at the outer region of the SMBH in the MW is described by the gNFW profile, whereas within the gravitational influence radius $r_{\text{sp}}$ it follows a spike distribution
\begin{equation}
\label{p-spike_1}
\rho_{\text{spike}}(r)=\rho_{\text{sp}}\left(1 - \frac{4R_\text{s}}{r}\right)^3 \left(\frac{r_{\text{sp}}}{r}\right)^{\gamma_{\mathrm{sp}}}.
\end{equation}
Here, $\rho_{\text{sp}}$ denotes a density at a reference radius $r_{\text{sp}}$, $R_\text{s}=2G m_1/{c^2}$ is the Schwarzschild radius of the SMBH with mass $m_1$ and 
\begin{align}
\label{gamma-spike}
\gamma_{\mathrm{sp}}=(9-2\gamma )/(4-\gamma ).
\end{align}
To sum up, the final density profile of a halo is composed of three layers

\begin{align}
\label{p-spike}
\rho(r) = 
\begin{cases}
0,& r \leq r_{\text{ISCO}},
\\ \rho_{\text{sp}}\left(1 - \frac{4R_\text{s}}{r}\right)^3 \left(\frac{r_{\text{sp}}}{r}\right)^{\gamma_{\mathrm{sp}}}, & r_{\text{ISCO}} < r \leq r_{\text{sp}}, \\
\rho_{\rm gNFW}(r),&r_{\text{sp}}<r.
\end{cases}
\end{align}

Next, we need to determine the two parameters of the DM spike  distribution, $r_{\text{sp}}$ and 
$\rho_{\text{sp}}$. By ensuring that the DM densities of inner spike profile and the outer gNFW profile are equal at 
$r=r_{\text{sp}}$
 , we can use this matching condition to derive 
 $\rho_{\text{sp}}= \rho_{\rm gNFW}(r_{\text{sp}})/(1-4R_s/r_{\text{sp}})^{-3}$. Here we assume that the distribution of DM around the BH initially follows a gNFW profile. The adiabatic growth of the BH produces a dense spike in the inner region of the minihalo within a radius of $r_{\text{sp}}$. The gravitational influence radius can be denoted as 
$4\pi \int_{0}^{r_{h} } \rho_{\rm gNFW} \left ( r \right ) r^{2}{{\dm}r}=2m_1 $~\cite{Merritt:2003qc}. Taking into account the influence radius of gravity and the constraints imposed by the matching condition at $r_{\text{sp}}$, we obtain the following system of equations
\begin{equation}\label{eq_spike_density}
\rho_{\text{sp}} = \frac{\rho_{\rm gNFW}(r_{\text{sp}})}{(1-4R_s/r_{\text{sp}})^{3}},
\end{equation}
and

\begin{equation}\label{eq_mass_integral}
 4\pi \int_{0}^{r_{h}} \rho_{\rm gNFW} \left( r \right) r^{2} {{\dm}r}=2m_1.
\end{equation}
For
$\gamma=0.5, 1, 1.5, 2$ in Eq.~\eqref{eq::gnfw_halo}, we list the relevant calculation results in Table~\ref{tab:2}.

\begin{table}[]
\centering

\begin{tabular}{ccccc}
\hline\hline
\(\gamma\) & 0.5    & 1      & 1.5    & 2      \\ \hline
\(\gamma_{\text{sp}}\) & 16/7   & 7/3    & 2.4    & 2.5    \\ \hline
\(r_{\text{sp}}\) (pc) & 27.309 & 12.664 & 3.289  & 0.192  \\ \hline
\(\rho_{\text{sp}}\) (\(M_{\odot}/\text{pc}^3\)) & 6.216  & 13.425 & \(1.705 \times 10^3\) & \(1.929 \times 10^7\) \\ \hline\hline
\end{tabular}
\caption{In the gNFW profile in Eq.~\eqref{eq::gnfw_halo} with \(\gamma = \{0.5, 1, 1.5, 2\}\), the adiabatic growth of the central BH leads to the formation of a DM spike in Eq.~\eqref{p-spike_1}. The corresponding values of \(\gamma_{\text{sp}} = \{16/7, 7/3, 2.4, 2.5\}\), \(r_{\text{sp}}\), and \(\rho_{\text{sp}}\) after the formation of the DM spike  are recorded in the table.}
\label{tab:2}
\end{table}

\begin{figure}
    \centering
    \includegraphics[height=0.2\textheight]{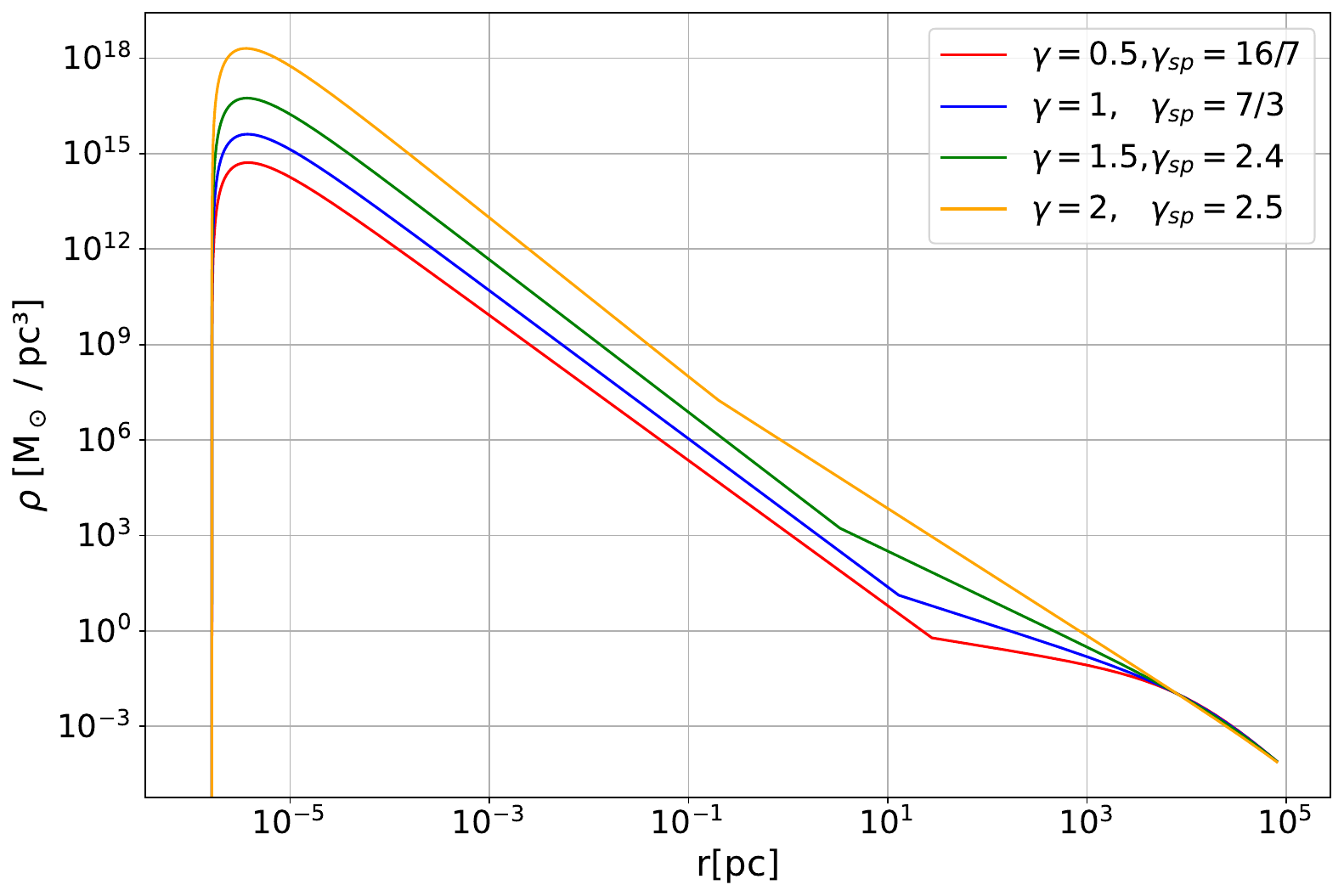}
    \caption{In the gNFW profile in Eq.~\eqref{eq::gnfw_halo}, the power-law indices \(\gamma = \{0.5, 1, 1.5, 2\}\) correspond to the formation of spike power-law indices \(\gamma_{\text{sp}} = \{16/7, 7/3, 2.4, 2.5\}\) in Eq.~\eqref{p-spike}. The relevant parameters are listed in Table~\ref{tab:2}.}
    \label{fig:spike}
\end{figure}
In Fig.~\ref{fig:spike}, we observe the formation of spike distributions under different initial power-law indices $\gamma$. Specifically, when the initial distribution follows an NFW profile in Eq.~\eqref{eq:nfw_halo} with \(\gamma = 1\), the resulting spike exhibits a power-law index of \(\gamma_{\text{sp}} = 7/3\). For the initial power-law index in the range \(1 \leq \gamma \leq 2\), the corresponding spike indices fall within \(2.25 \leq \gamma_{\text{sp}} \leq 2.5\). Comparing with the gNFW profiles in Figure~\ref{fig:gNFW_profile}, we note that in the absence of a spike, the DM density profiles near the GC exhibit significant discrepancies for different power-law indices. However, after the formation of a spike, the slope and density variations become more pronounced for distributions with smaller initial \(\gamma\) values. Furthermore, the differences in DM density near the GC among spikes with different power-law indices are considerably smaller compared to the case without a spike.

\section{Perturbed Kepler Orbits}
\label{sec:per}

\begin{figure}
	\centering
	\includegraphics[height=0.25\textheight]{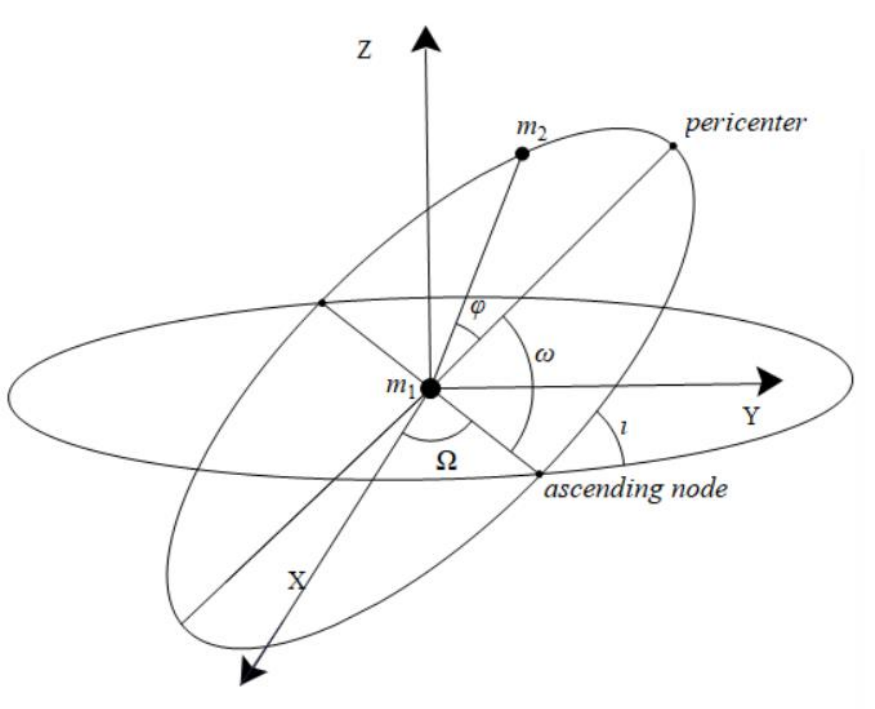}
	\caption{The schematic diagram of orbital motion viewed in the fundamental reference frame.}
	\label{fig:orbit}
\end{figure}

In this section, we present the method of osculating orbits.
The BBH system is composed of a primary mass, denoted as \( m_1 \), and a secondary mass, \( m_2 \), both idealized as Schwarzschild BHs for simplicity, as illustrated in Fig.~\ref{fig:orbit}. Within this framework, we consider a scenario where a smaller BH undergoes Keplerian motion around the SMBH located at the center of the MW. The Keplerian orbit is described by the following equation
\begin{equation} \label{kepler_orbit}
    r = \frac{p}{1 + e \cos(\varphi)},
\end{equation}
where \( r \) is the radial distance, \( p \) is the semi-latus rectum, \( e \) is the orbital eccentricity, and \( \varphi \) is the true anomaly. The radial and angular velocities of a Keplerian orbit are given by
\begin{align} \label{kepler_r_v}
    \dot{r} &= \sqrt{\frac{Gm}{p}} \, e \sin(\varphi),\\
\label{kepler_phi_v}
    \dot{\varphi} & = \sqrt{\frac{Gm}{p^3}} \, \left[1 + e \cos(\varphi)\right]^2,
\end{align}
where \( m := m_1 + m_2 \) is the total mass of the system.

 While exact analytical solutions do not exist for the three-body or many-body problems, in certain cases where the gravitational influence of additional bodies is sufficiently weak, the system can be treated as a perturbed two-body problem. The relative acceleration of two bodies in a Keplerian orbit is given by~\cite{poisson2014gravity}
\begin{equation} \label{a-k}
    \bm{a} = -\frac{Gm}{r^2} \bm{n} + \bm{f},
\end{equation}
where \( \bm{r} := \bm{r}_1 - \bm{r}_2 \) is the relative position vector, \( \bm{n} := \bm{r}/r \) is the unit radial vector, and \( \bm{f} \) represents the perturbing force per unit mass. The perturbing force can be expressed as
\begin{equation} \label{p-f}
    \bm{f} = \mathcal{R} \, \bm{n} + \mathcal{S} \, \bm{k} + \mathcal{W} \, \bm{e}_z,
\end{equation}
where \( \bm{k} \) is the unit vector orthogonal to \( \bm{n} \), and \( \bm{e}_z \) is the normal vector to the orbital plane.

The equations governing the osculating orbital elements are given by
\begin{align}
	\label{dp-dt}
	\frac{{\dm}p}{{\dm} t} &= 2\sqrt{\frac{p^3}{Gm}} \frac{1}{1+e\cos\varphi} \mathcal{S}, \\
	\label{de-dt}
	\frac{{\dm}e}{{\dm} t} &= \sqrt{\frac{p}{Gm}} \left[\sin\varphi \, \mathcal{R} + \frac{2\cos\varphi + e(1+\cos^2 \varphi)}{1+e\cos\varphi} \mathcal{S}\right], \\
	\label{dphi-dt}
	\frac{{\dm}\varphi}{{\dm} t} &= \sqrt{\frac{G m}{p^3}} \left(1+e\cos\varphi\right)^2 \nonumber \\
	&\quad + \frac{1}{e}\sqrt{\frac{p}{G m}} \left[\cos\varphi \, \mathcal{R} - \frac{2+e\cos\varphi}{1+e\cos\varphi} \sin\varphi \, \mathcal{S}\right].
\end{align}

Here, we assume that the orbital elements \(\omega\) and \(\iota\) are both zero. When the perturbative force is very small, such that the changes in the orbital elements are minimal, it is convenient to use $\varphi$ as the independent variable instead of $t$. 
Under this approximation, the non-Keplerian terms on the right-hand side of the equations can be neglected, allowing for a good approximation of the orbital dynamics. By combining Eqs.~\eqref{dp-dt},Eq.~\eqref{de-dt} andEq.~\eqref{dphi-dt}, and expanding to the first-order term in \(p^2/(Gm)\), we obtain
\begin{align}
	\label{dp-dphi}
	\frac{{\dm}p}{{\dm}\varphi} &= 2\frac{p^3}{Gm} \frac{1}{(1+e\cos\varphi)^3} \mathcal{S}, \\
	\label{de-dphi}
	\frac{{\dm}e}{{\dm}\varphi} &= \frac{p^2}{Gm} \left[\frac{\sin\varphi}{(1+e\cos\varphi)^2} \mathcal{R} \right. \nonumber \\
	&\quad \left. + \frac{2\cos\varphi + e(1+\cos^2\varphi)}{(1+e\cos\varphi)^3} \mathcal{S}\right], \\
	\label{dt-dphi}
	\frac{{\dm} t}{{\dm}\varphi} &= \sqrt{\frac{p^3}{G m}} \frac{1}{(1+e\cos\varphi)^2} \left\{1 - \frac{1}{e}\frac{p^2}{G m} \right. \nonumber \\
	&\quad \left. \times \left[ \frac{\cos\varphi}{(1+e\cos\varphi)^2} \mathcal{R} - \frac{(2+e\cos\varphi)\sin\varphi}{(1+e\cos\varphi)^3} \mathcal{S} \right] \right\}.
\end{align}
For Eq.~\eqref{dt-dphi}, since \(p^2/(Gm) \ll 1\), the contribution of the second term can be neglected, yielding
\begin{equation} \label{dt_dphi}
\frac{{\dm} t}{{\dm}\varphi} = \sqrt{\frac{p^3}{G m}} \frac{1}{(1+e\cos\varphi)^2}.
\end{equation}

The perturbing force per unit mass \(\bm{f}\), induces both oscillations and secular changes in the orbital elements. To quantify the cumulative drift associated with these secular effects, we average over a complete orbital period. The orbital average of \(\left<\dot{a} \right>\) is defined as
\begin{equation} \label{average}
\left<\frac{{\dm}a}{{\dm}t}\right> = \frac{1}{T} \int_{0}^{P} \frac{{\dm}a}{{\dm}t} \, {{\dm}t}= \frac{1}{T} \int_{0}^{2\pi} \frac{{\dm}a}{{\dm}\varphi} \, {{\dm}\varphi},
\end{equation}
where \(T\) denotes the orbital period.

\subsection{Reaction of Gravitational Waves}

GWs carry away the orbital energy of binary systems, acting as a reaction force that diminishes the orbital eccentricity and drives the orbits closer together. This reaction can be treated as a perturbative force, and its effect on the acceleration of the system is given by ~\cite{poisson2014gravity}

\begin{align} 
\label{a-gw-rr}
	\begin{split}
		\bm{a}_{\text{GW}} =& \frac{8}{5} \frac{G^2 m \mu}{c^5 r^3}\left[\left(3v^2 + \frac{17}{3}\frac{G m}{r}\right)\dot{r}\bm{n}\right.\\
		&\qquad\qquad \left.-\left(v^2 + 3\frac{G m}{r}\right)\bm{v}\right],
	\end{split}
\end{align}
 where $\mu=m_1m_2/(m_1+m_2)$ is the reduced mass.
 
At the lowest order in a post-Newtonian expansion and using the quadrupole formula, the standard results yield the expressions for the secular changes in the semi-latus rectum and eccentricity $e$ due to GW emission~\cite{poisson2014gravity,Cardoso:2020iji}

\begin{align}
	\label{pk-rr-pf}
	\left\langle\frac{{\dm}p}{{\dm} t}\right\rangle_{\text{GW}} &= -\frac{8}{5}\eta\frac{(G m)^{3}}{c^5 p^{3}}\left(1-e^2\right)^{3/2}\left(8 + 7e^2\right), \\
	\label{pk-rr-ef}
	\left\langle\frac{{\dm}e}{{\dm} t}\right\rangle_{\text{GW}} &= -\frac{8}{5}\eta\frac{(G m)^{3}}{c^5 p^{4}}\left(1-e^2\right)^{3/2}\left(\frac{304}{24}e + \frac{121}{24}e^3\right),
\end{align}
where \(\eta = m_1m_2/(m_1 + m_2)^2\) is the symmetric mass ratio, and the subscript ``GW" indicates that these effects arise from the reaction of GWs. From Eqs.~\eqref{pk-rr-pf} and \eqref{pk-rr-ef}, it is evident that GW emission leads to a gradual reduction in both the semi-latus rectu and the eccentricity of the binary system.

\subsection{Dynamical Friction and Accretion}
\label{DFA}

Chandrasekhar~\cite{Chandrasekhar:1943ys} proposed that an object moving through an infinite, homogeneous medium experiences a drag force due to gravitational interactions, acting in the direction opposite to its velocity. This effect, known as dynamical friction, depends on the velocity of the moving object, as well as the density and sound speed of the surrounding medium~\cite{Ostriker:1998fa, Kim:2007zb}.
The SMBH at the center of the MW is surrounded by a DM halo. In this scenario, we consider a small BH orbiting the SMBH while moving through the DM medium. As a result, the small BH will experience dynamical friction.
Without loss of generality, we focus on the supersonic regime, where the dynamical friction force can be expressed as
\begin{equation} \label{f-df}
	\bm{f}_{\text{DF}} = -\frac{4\pi G^2 {m_2}^2 \rho_{\text{DM}} I_{v}}{v^3} \bm{v},
\end{equation}
where \(\bm{v}\) is the velocity of the small BH, which can be decomposed into \(\bm{v} = \dot{r}\bm{n} + r\dot{\varphi}\bm{k}\). Here, \(I_v\) is the Coulomb logarithm $I_v=\left \{ 3,10,\log{\sqrt{\frac{m_1}{m_2} } }  \right \} $~\cite{Eda:2014kra, Yue:2019ozq, Kavanagh:2020cfn}, which depends on the velocity \(v\) and the sound speed of the DM halo. In this work, we adopt \(I_v = \log\left(\sqrt{\frac{m_1}{m_2}}\right)\). 

Eq.~\eqref{f-df} can be rewritten as
\begin{equation} \label{dp_dt_df}
	\bm{f}_{\text{DF}} = -\frac{4\pi G^2 {m_2}^2 \rho_{\text{DM}} I_{v}}{v^3} \left(\dot{r}\bm{n} + r\dot{\varphi}\bm{k}\right).
\end{equation}
Substituting the above equation into Eqs.~\eqref{dp-dphi},  \eqref{de-dphi}, and combining with Eq.~\eqref{average}, we obtain the following equations:
\begin{align} \label{dp_dt_df_avg}
	\begin{split}
	\left\langle\frac{{\dm}p}{{\dm} t}\right\rangle_{\text{DF}} &= -\frac{1}{T} \int_{0}^{2\pi} 2\frac{p^3}{Gm} \frac{1}{(1+e\cos\varphi)^3} \\
	&\quad \times \left( \frac{4\pi G^2 {m_2} \rho_{\text{DM}} I_{v}}{v^3} r\dot{\varphi} \right) {{\dm}\varphi}\\
    &=- \int_{0}^{2\pi}\frac{4G^{1/2}{m_2} \rho_{DM} I_{v} p^{5/2}}{(1-e^2)^{-3/2}m^{3/2}}f(\varphi){{\dm}\varphi},
	\end{split}
\end{align}
where $f(\varphi)=\frac{1}{(1+e\cos\varphi)^{2}(e^2+2e\cos\varphi+1)^{3/2}}$,
\begin{align}\label{de_dt_df_avg}
	\begin{split}
	\left\langle\frac{{\dm}e}{{\dm} t}\right\rangle_{\text{DF}} &= - \frac{1}{T} \int_{0}^{2\pi} \frac{p^2}{Gm} \times \\
&\left[\frac{2\cos\varphi+e(1+\cos^2 \varphi)}{(1+e\cos\varphi)^3} \left( \frac{4\pi G^2 {m_2} \rho_{\text{DM}} I{_v}}{v^3}r\dot{\varphi} \right)
 \right. \\
    & \qquad\qquad \left. \hspace{-0.5cm} 
+    	\frac{\sin\varphi}{(1+e\cos\varphi)^2} \left( \frac{4\pi G^2{m_2} \rho_{\text{DM}} I_{v}}{v^3}\dot{r} \right)\right] {{\dm}\varphi}\\
&=-\int_{0}^{2\pi} \frac{4p^{3/2}G^{1/2}{m_2} \rho_{DM} I_{v}}{(1-e^2)^{-3/2}m^{3/2}}g(\varphi){{\dm}\varphi},
	\end{split}
\end{align}
where $g(\varphi)=\frac{e+\cos\varphi}{(1+e\cos\varphi)^{2}(e^2+2e\cos\varphi+1)^{3/2}}$.

We now turn to the discussion of accretion. As the small BH moves through the DM environment, it will accrete DM. 
The accretion of DM can lead to significant dephasing, which generally cannot be neglected~\cite{Karydas:2024fcn}.
In our analysis, we assume that the mass of the small BH does not exceed $1\%$ of the mass of central SMBH. This allows us to neglect the disruption of the DM distribution caused by the accretion of the small BH.
In this study, we assume that the radius of the small BH exceeds the mean free path of DM particles. Furthermore, we focus exclusively on non-annihilating DM particles, neglecting all interactions other than gravitational effects. Under these conditions, the accretion process of the small BH is described by Bondi-Hoyle accretion. Consequently, the mass flux at the event horizon of the small BH is given by ~\cite{Macedo:2013qea,Mach:2021zqe}
\begin{equation} \label{ut-dm}
	\dot{m_2} = 4\pi G^2 \lambda \frac{{m_2}^2 \rho_{\text{DM}}}{(v^2 + c_s^2)^{3/2}},
\end{equation}
where \(\lambda\) is a dimensionless parameter of order unity, dependent on the properties of the DM medium, and \(c_s\) denotes the sound speed of the DM medium. For simplicity, we assume \(v \gg c_s\) and adopt \(\lambda = 1\) throughout this work.

Due to the accretion of DM, the equation of motion can be written as
\begin{equation} \label{eom-a-dm}
    m_2\dot{\bm{v}} + \dot{m}_2\bm{v} = -\frac{G m_1 m_2}{r^3}\bm{r},
\end{equation}
where the force arising from the accretion term $\dot{m}_2 \bm{v}$ is given by
\begin{equation}\label{f-aa}
    \bm{f}_{\text{AC}} = -\frac{4\pi G^2 m_2^2 \rho_{\text{DM}} \lambda}{v^3}\bm{v}.
\end{equation}
Similar to dynamical friction, we can obtain the following equations:
\begin{align}\label{dp_dt_ac}
\left\langle\frac{{\dm}p}{{\dm} t}\right\rangle_{\text{AC}} = - \int_{0}^{2\pi}\frac{4G^{1/2}{m_2} \rho_{DM}\lambda p^{5/2}}{(1-e^2)^{-3/2}m^{3/2}}f(\varphi){{\dm}\varphi},
\end{align}

\begin{align}\label{de_dt_ac}
\left\langle\frac{{\dm}e}{{\dm} t}\right\rangle_{\text{AC}} = -\int_{0}^{2\pi} \frac{4p^{3/2}G^{1/2}{m_2} \rho_{DM}\lambda}{(1-e^2)^{-3/2}m^{3/2}}g(\varphi){{\dm}\varphi}.
\end{align}
\begin{figure}
\centering
\includegraphics[width=1\linewidth, height=0.25\textheight]{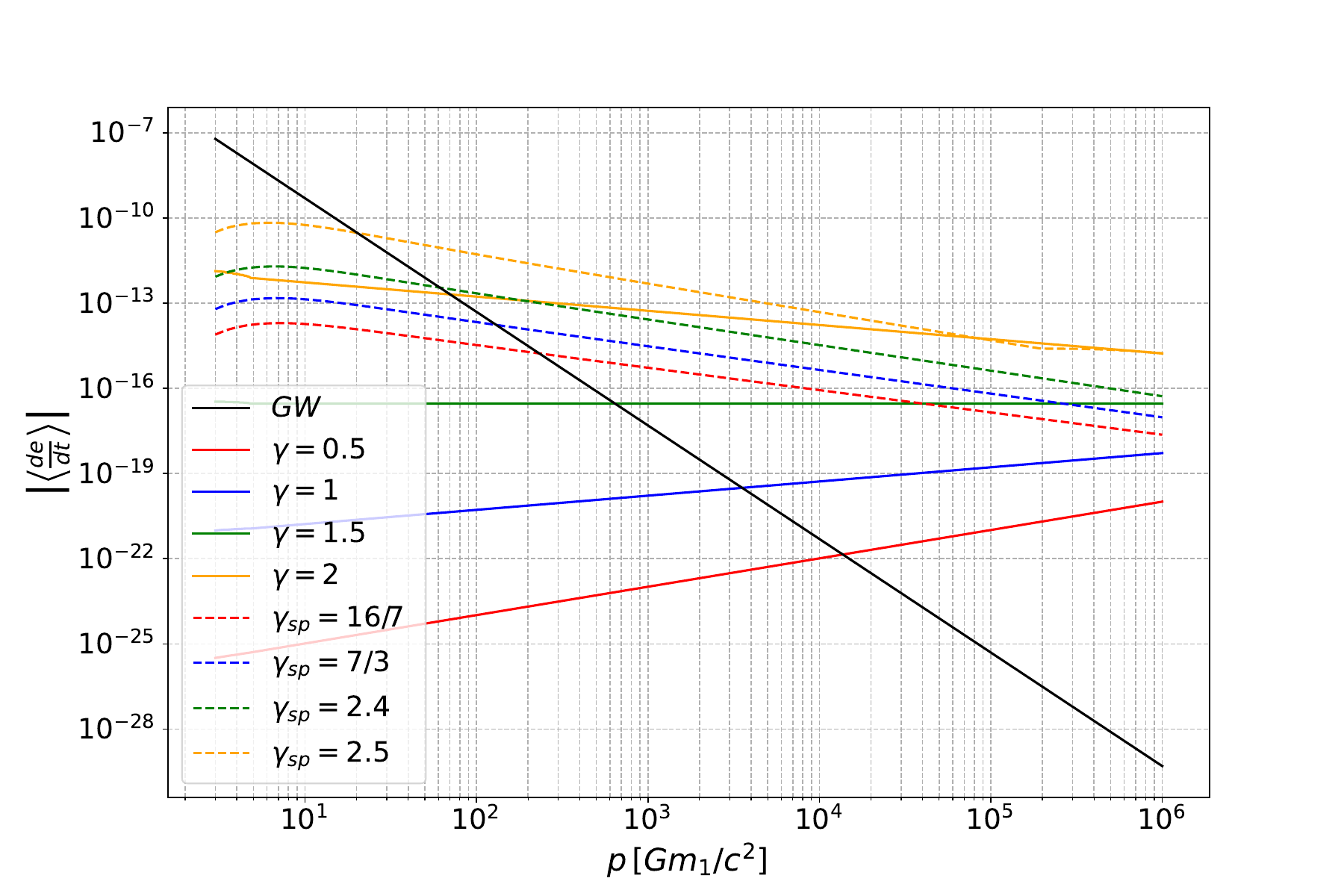}
\caption{The variation of $\left|\left\langle\frac{de}{{\dm}t}\right\rangle\right|$ with respect to the initial semi-latus rectum $p=p_0$ is analyzed, where we assume $e = 0.6$ and $m_2 = 1000\, M_{\odot}$. The solid black line represents the eccentricity change due to GWs, with $\left|\left\langle\frac{de}{{\dm}t}\right\rangle_{\text{GW}}\right|$ in Eq.~\eqref{pk-rr-ef}. The solid line corresponds to the rate of eccentricity change 	$\left|\left\langle\frac{{\dm}e}{{\dm} t}\right\rangle_{\text{AC}}\right|$ in Eq.~\eqref{de_dt_ac} for the gNFW profile in Eq.~\eqref{eq::gnfw_halo}, and the dashed line represents the case for a DM spike in Eq.~\eqref{p-spike_1}.}
\label{fig:de_dphi_1}
\end{figure}

From Eqs.~\eqref{pk-rr-pf} and \eqref{pk-rr-ef}, we see that gravitational radiation decreases both the eccentricity $e$ and the orbital semi-latus rectum $p$. For dynamical friction and accretion, as shown in Eqs.~\eqref{f-df} and \eqref{f-aa}, they share the same functional form. Through Eqs.~\eqref{dp_dt_df_avg} and \eqref{dp_dt_ac}, we observe that they reduce the semi-latus rectum $p$ and accelerate the merger. However, for the eccentricity evolution, since it involves the term $g(\varphi)=\frac{e+\cos\varphi}{(1+e\cos\varphi)^{2}(e^2+2e\cos\varphi+1)^{3/2}}$, the numerator in the integral can become negative, potentially leading to an increase in eccentricity. Given that the forces generated by dynamical friction and accretion have identical structures, we focus exclusively on studying the effect of accretion on eccentricity.

As illustrated in Fig.~\ref{fig:de_dphi_1}, we investigate the rate of eccentricity change $\left\langle\frac{de}{{\dm}t}\right\rangle$, for an initial eccentricity $e_0 = 0.6$, using the initial semi-latus rectum $p_0$ as the independent variable to examine how $\left\langle\frac{de}{{\dm}t}\right\rangle$ varies with $p_0$. We compare $\left|\left\langle\frac{de}{{\dm}t}\right\rangle_{\text{GW}}\right|$ and $\left|\left\langle\frac{de}{{\dm}t}\right\rangle_{\text{AC}}\right|$, where the absolute values are taken because $\left\langle\frac{de}{{\dm}t}\right\rangle_{\text{GW}}$ is negative. Our findings reveal that at larger BBH separations, the rate of eccentricity change is primarily dominated by accretion, with the effect becoming more pronounced as the DM density increases. For gNFW DM distributions in Eq.~\eqref{eq::gnfw_halo}, significant differences arise depending on the power-law index. For $\gamma = 2$, the accretion begins to dominate over GW radiation at $p_0 \approx 100\, Gm_1/c^2$. For $\gamma = 0.5$, accretion reaches a magnitude comparable to GW radiation at $p_0 \approx 10000\, Gm_1/c^2$. When DM forms a spike, its effect becomes significant around $p_0 \approx 100\, Gm_1/c^2$.

\subsection{Total Effect of Dark Matter}
\label{near range}

In this subsection, we comprehensively consider the effects of dynamical friction, accretion, and GW radiation reaction as perturbations on the eccentricity and semi-latus rectum of the small BH. Here, we neglect the influence of the DM gravitational potential, as well as the feedback effects arising from dynamical friction and accretion.
The force per unit mass acting on the small BH is given by

\begin{equation}
	\label{a_tot}
	\bar{\bm{a}} _{\text{tot}}=\bm{a}_{\text{DF}}+\bm{a}_{\text{AC}}+\bm{a}_{\text{GW}}.
\end{equation}
Since the three forces mentioned above are treated as perturbations, we still approximate the motion of the secondary object as following an elliptical orbit. The rate of change of $\varphi$ with respect to time is given by
\begin{equation} 
\label{dphi_dt}
\frac{{\dm}\varphi}{{\dm}t} = \sqrt{\frac{G m}{p^3}} (1 + e \cos \varphi)^2.
\end{equation}
 The general expressions for the semi-latus rectum $p$ and eccentricity $e$ are as follows

\begin{equation}
	\label{dp_dt_tot}
	\left\langle\frac{{\dm}p}{{\dm} t}\right\rangle_{\text{tot}}=\left\langle\frac{{\dm}p}{{\dm} t}\right\rangle_{\text{DF}}+\left\langle\frac{{\dm}p}{{\dm} t}\right\rangle_{\text{AC}}+\left\langle\frac{{\dm}p}{{\dm} t}\right\rangle_{\text{GW}},
    \end{equation}
\begin{equation}    
	\label{de_dt_tot}
	\left\langle\frac{{\dm}e}{{\dm} t}\right\rangle_{\text{tot}}=\left\langle\frac{{\dm}e}{{\dm} t}\right\rangle_{\text{DF}}+\left\langle\frac{{\dm}e}{{\dm} t}\right\rangle_{\text{AC}}+\left\langle\frac{{\dm}e}{{\dm} t}\right\rangle_{\text{GW}}.
    \end{equation}

\begin{figure*}
    \centering  \includegraphics[width=1\linewidth, height=0.35\textheight]{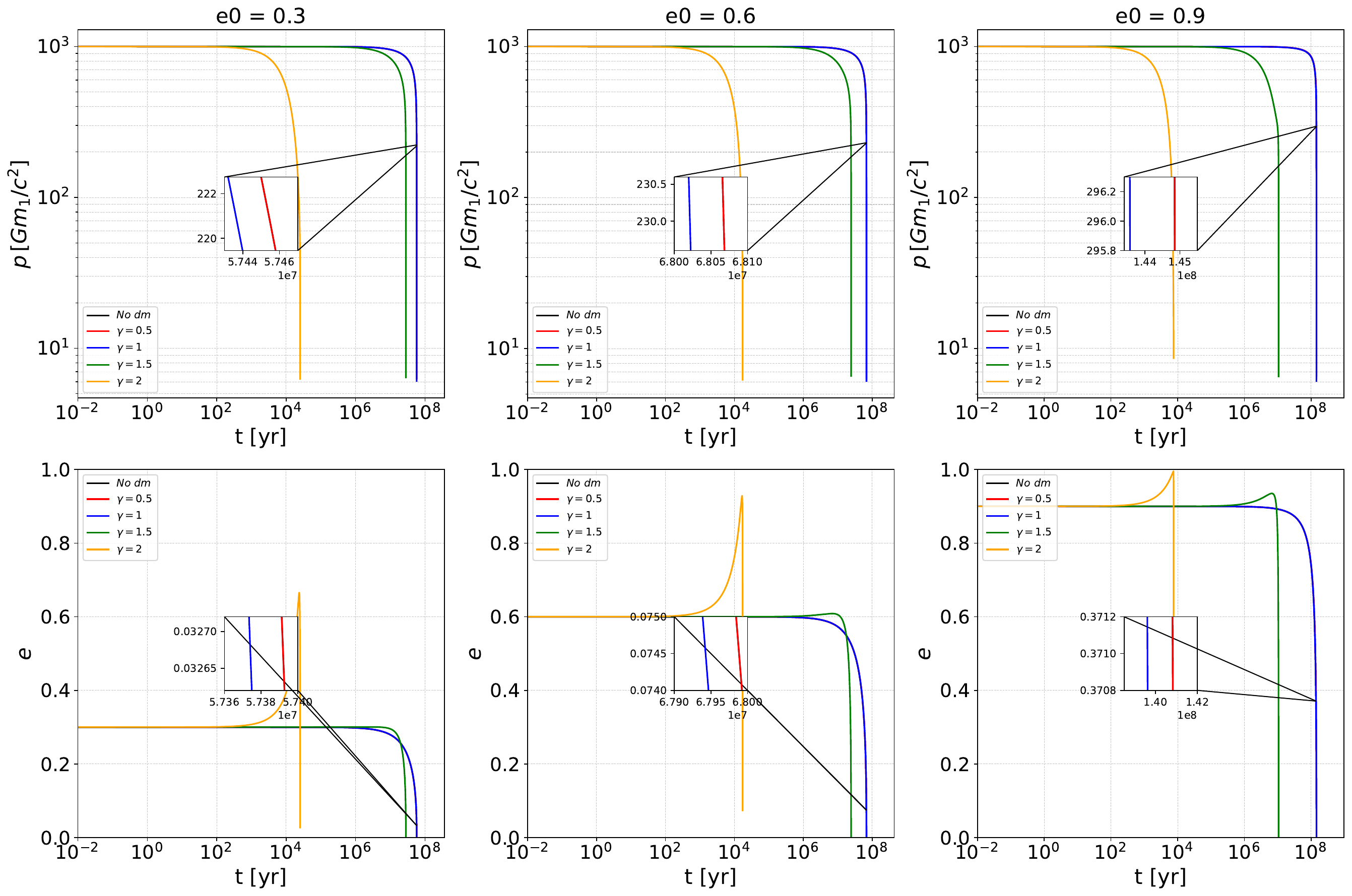}
   \caption{This figure presents the temporal evolution of eccentricity and semi-latus rectum $p$ for BBH systems embedded in a gNFW profile of  DM distribution in Eq.~\eqref{eq::gnfw_halo} with slope $\gamma = \{0.5, 1, 1.5, 2\}$. The initial eccentricities $e_0$ are $\{0.3, 0.6, 0.9\}$ for each case. The top row shows the evolution of the semi-latus rectum $p$ with time, while the bottom row displays the corresponding eccentricity evolution. In all panels, the black curves represent the reference case without DM. For the plots of semi-latus rectum $p$ (top row), the horizontal axis shows time in years (yr), and the vertical axis is normalized in units of $Gm_1/c^2$. The initial semi-latus rectum is set to $p_0=1000\,Gm_1/c^2$, where $m_1 = 4.26 \times 10^6\,M_\odot$ is the primary BH mass and the secondary BH has a mass of $m_0=1000\,M_\odot$.  The evolution of semi-latus rectum $p$ and eccentricity $e$ terminates when $r < r_{\mathrm{ISCO}}$.}
    \label{fig:gNFW_p_one}
\end{figure*}

Taking into account the combined effects of dynamical friction, accretion, and GW radiation, we have plotted the changes in eccentricity $e$ and semi-latus rectum $p$ of a small BH under the generalized gNFW  model. The evolution of semi-latus rectum $p$ and eccentricity $e$ terminates when $r \leq r_{\mathrm{ISCO}}$.
As shown in Fig.~\ref{fig:gNFW_p_one}, the red, blue, green, and yellow curves represent the temporal evolution of the semi-latus rectum $p$ and eccentricity $e$ under different DM density profiles characterized by $\gamma \in \{0.5, 1, 1.5, 2\}$, considering only the pure gNFW profile without accounting for potential DM spike formation near the BHs. 

The orbital evolution for systems with initial semi-latus rectum $p_0 = 1000\,G m_1/c^2$ $(\sim0.2\rm mpc)$ exhibits characteristic behavior, where in the absence of DM (black curves), the orbital decay proceeds exclusively through GW emission. When DM is included, the orbital decay rate increases significantly for steeper density profiles ($\gamma \geq 1.5$), with a clear separation visible between the $\gamma = 2$ and $\gamma = 1.5$ cases throughout their evolution. For shallower profiles ($\gamma \leq 1$), the effects become progressively weaker: while the $\gamma = 1$ case shows marginally detectable deviations upon close inspection, the $\gamma = 0.5$ profile remains practically indistinguishable from the DM free scenario. 

The eccentricity evolution shown in the lower panel of Fig.~\ref{fig:gNFW_p_one} demonstrates a similar dependence on the 
$\gamma$ parameter. The steep $\gamma = 2$ profile produces significant eccentricity enhancement, while $\gamma = 1.5$ shows a modest but discernible increase. Profiles with $\gamma \leq 1$ demonstrate no measurable eccentricity modification compared to the vacuum case. These results collectively indicate that dynamical friction effects scale monotonically with the central density slope, where only sufficiently steep gNFW profiles ($\gamma \geq 1.5$) generate observable perturbations to both the orbital decay rate and eccentricity evolution, while shallower profiles ($\gamma \leq 1$) remain observationally degenerate with the pure GW emission scenario at current detection thresholds.

\begin{figure*}
    \centering
    \includegraphics[width=1\linewidth, height=0.35\textheight]{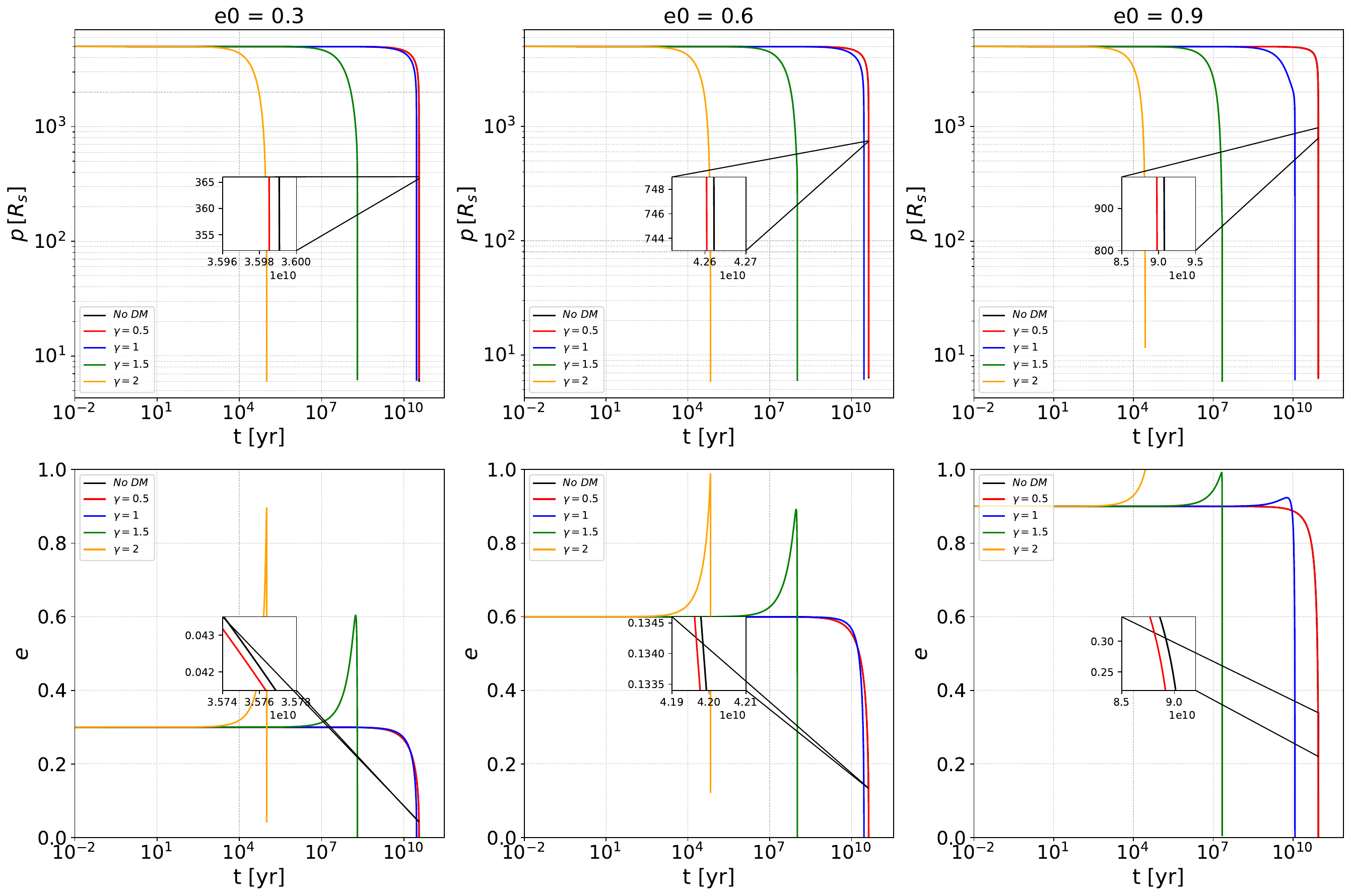}
    \caption{Orbital evolution for a binary system with a secondary BH mass of $10^3 M_\odot$ and initial semi-latus rectum $p_0 = 5000\,Gm_1/c^2$. The top row shows the temporal evolution of the semi-latus rectum $p$ for three different initial eccentricities: $e_0 \in \{0.3, 0.6, 0.9\}$. The bottom row displays the corresponding eccentricity evolution for the same set of initial conditions, illustrating how both orbital parameters evolve differently depending on the initial eccentricity configuration.}
    \label{fig:gNFW_p_five}
    
\end{figure*}

For an initial semi-latus rectum of $p_0 = 5000\,Gm_1/c^2 (\sim1\rm mpc)$, Fig.~\ref{fig:gNFW_p_five} reveals several key features of the orbital evolution. The top panels demonstrate that the orbital decay rates for $\gamma=1$  and $\gamma=0.5$  become distinguishable, while the $\gamma=0.5$ case initially appears degenerate with the dark-matter-free scenario. However, detailed inspection of magnified regions confirms their subtle but measurable separation. Notably, this wider binary separation ($p_0 = 5000\,Gm_1/c^2$) provides better discrimination between different DM profile slopes $\gamma$ compared to the $1000\,Gm_1/c^2$ case.
The eccentricity evolution exhibits more pronounced DM effects at larger orbital scales. Compared to the case of $p_0 = 1000 \, G m_1 / c^2$, the $\gamma = 2$ density profile produces significantly stronger eccentricity enhancement, while $\gamma=1.5$ shows substantial increase and $\gamma=1$ displays modest but detectable growth. This systematic dependence on both $\gamma$ and initial orbital separation suggests that wider binaries are more sensitive probes of the DM density profile's inner slope.

As evident from Fig.~\ref{fig:gNFW_p_one}, the yellow and blue curves exhibit close proximity during their initial descent, while their separation becomes increasingly pronounced with higher initial eccentricity. In Fig.~\ref{fig:gNFW_p_five}, we also observe that the blue and red curves remain close to each other at small eccentricities, but gradually diverge as the initial eccentricity increases. For the initial semi-latus rectum $p_0 = 1000 \, Gm_1/c^2$, we observe that the merger times for $\gamma=2$ and $\gamma=1.5$ decrease with increasing initial eccentricity, while the blue curve in the figure shows an increase in merger time. When the initial semi-latus rectum is $p_0 = 5000 \, Gm_1/c^2$, the figure reveals that the merger times for $\gamma=2$, $\gamma=1.5$, and $\gamma=1$ all decrease with growing initial eccentricity, whereas for $\gamma=0.5$ the merger time increases. Due to the termination of evolution when $r < r_{\mathrm{ISCO}}$, from Fig.~\ref{fig:gNFW_p_one} and Fig.~\ref{fig:gNFW_p_five},  we can observe  that in the case of $\gamma=2$, the eccentricity $e$ has not yet circularized before falling within $r_{\mathrm{ISCO}}$.

\begin{figure*}
    \centering
    \includegraphics[width=1\linewidth, height=0.35\textheight]{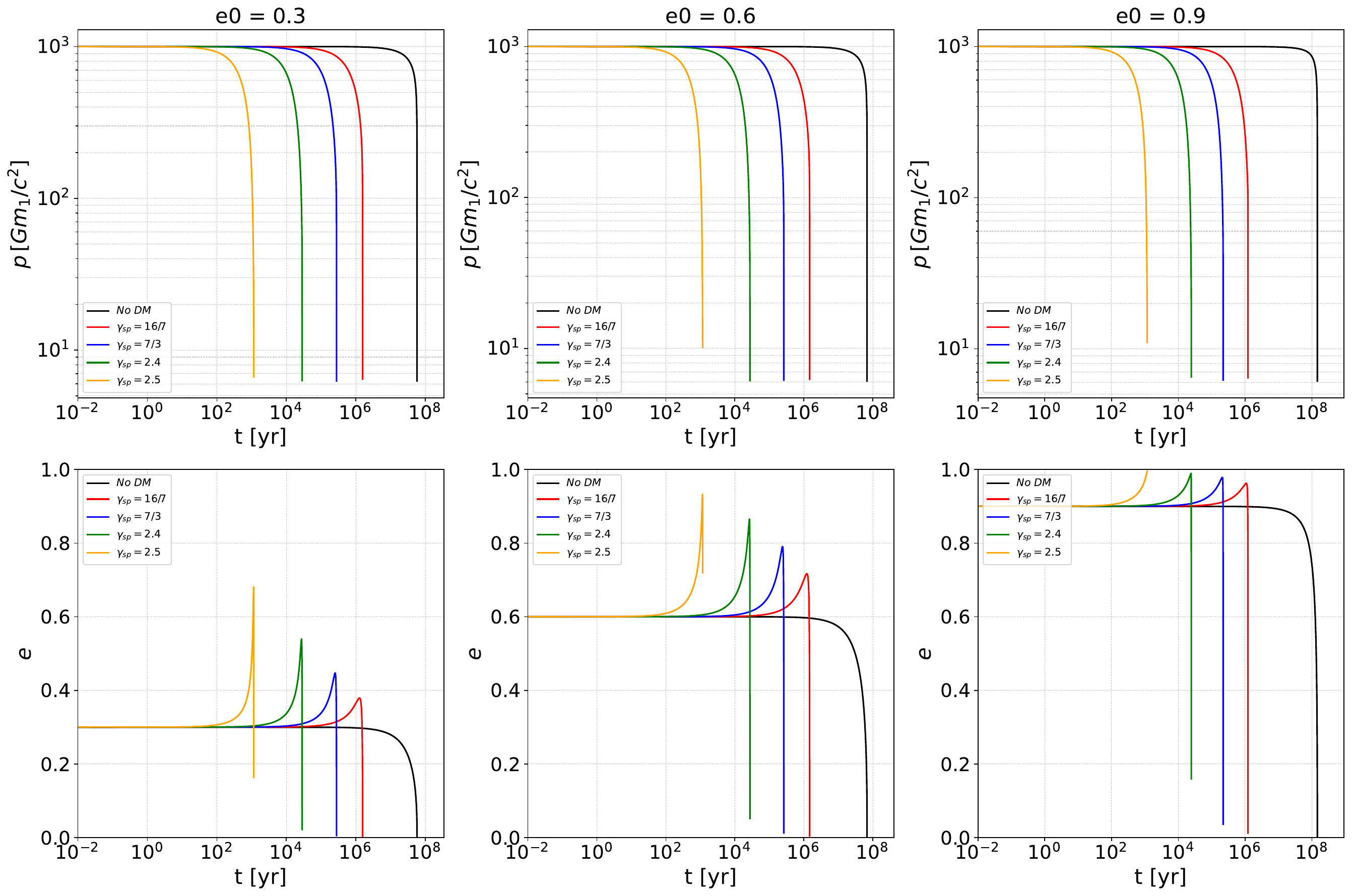}
    \caption{ Orbital evolution of a secondary BH in a DM spike environment with the profile in Eq. \eqref{p-spike_1}. 
        The DM density profile is modeled as a spike-modified gNFW profile in Eq. \eqref{p-spike}, 
        formed through adiabatic compression by the central SMBH. 
        \textbf{Top row}: Temporal evolution of the semi-latus rectum $p$ (in units of $Gm_1/c^2$) 
        with time in years (yr). \textbf{Bottom row}: Corresponding eccentricity evolution. 
        Initial conditions are fixed at $p_0 = 1000\,Gm_1/c^2$ with increasing initial 
        eccentricities (left to right: $e_0 = 0.3$, $0.6$, $0.9$).}
    \label{fig:p_spike_one}
\end{figure*}

Our results indicate a significant modification of the standard gNFW DM profile near the GC SMBH. Through adiabatic compression, the profile transforms into a steeper spike configuration with enhanced density distribution. The original gNFW power-law indices $\gamma\in\{0.5,1,1.5,2\}$ evolve into corresponding spike indices $\gamma_{\mathrm{sp}}\in\{16/7,7/3,2.4,2.5\}$.
Fig.~\ref{fig:p_spike_one} shows the orbital evolution of a secondary BH with an initial semi-latus rectum $p$ of $p_0=1000\,Gm_1/c^2$. Compared to the case without a DM spike (Fig.~\ref{fig:gNFW_p_one}), we observe that the formation of a DM spike significantly enhances both dynamical friction and accretion effects. In the absence of a spike, the profiles with $\gamma=1$ and 0.5 are nearly identical to the no-dark-matter case. When a DM spike forms, steeper profiles with higher $\gamma_{\mathrm{sp}}$ values exhibit more pronounced environmental effects, showing faster orbital semi-latus rectum decay in denser DM environments and more significant eccentricity enhancement compared to the no-spike case, where both effects intensify with increasing $\gamma_{\mathrm{sp}}$.

For an initial semi-latus rectum of $p_0=1000\,Gm_1/c^2$, we observe that different initial eccentricities $e_0=\{0.3,0.6,0.9\}$ have negligible impact on the total merger duration across various power-law profiles. The eccentricity evolution shows distinct patterns: when $e_0=0.3$, $\gamma_{\mathrm{sp}}=\{16/7,7/3,2.4,2.5\}$ produces $\Delta e\sim\{0.1,0.2,0.3,0.4\}$; for $e_0=0.6$, the maximum $\Delta e\sim\{0.1,0.2,0.25,0.3\}$; while $e_0=0.9$ yields only $\Delta e\sim0.1$. This clearly demonstrates that systems with lower initial eccentricities undergo more substantial relative changes in eccentricity during their evolution, with the effect diminishing as the initial eccentricity increases.

\begin{figure*}
    \centering
    \includegraphics[width=1\linewidth, height=0.35\textheight]{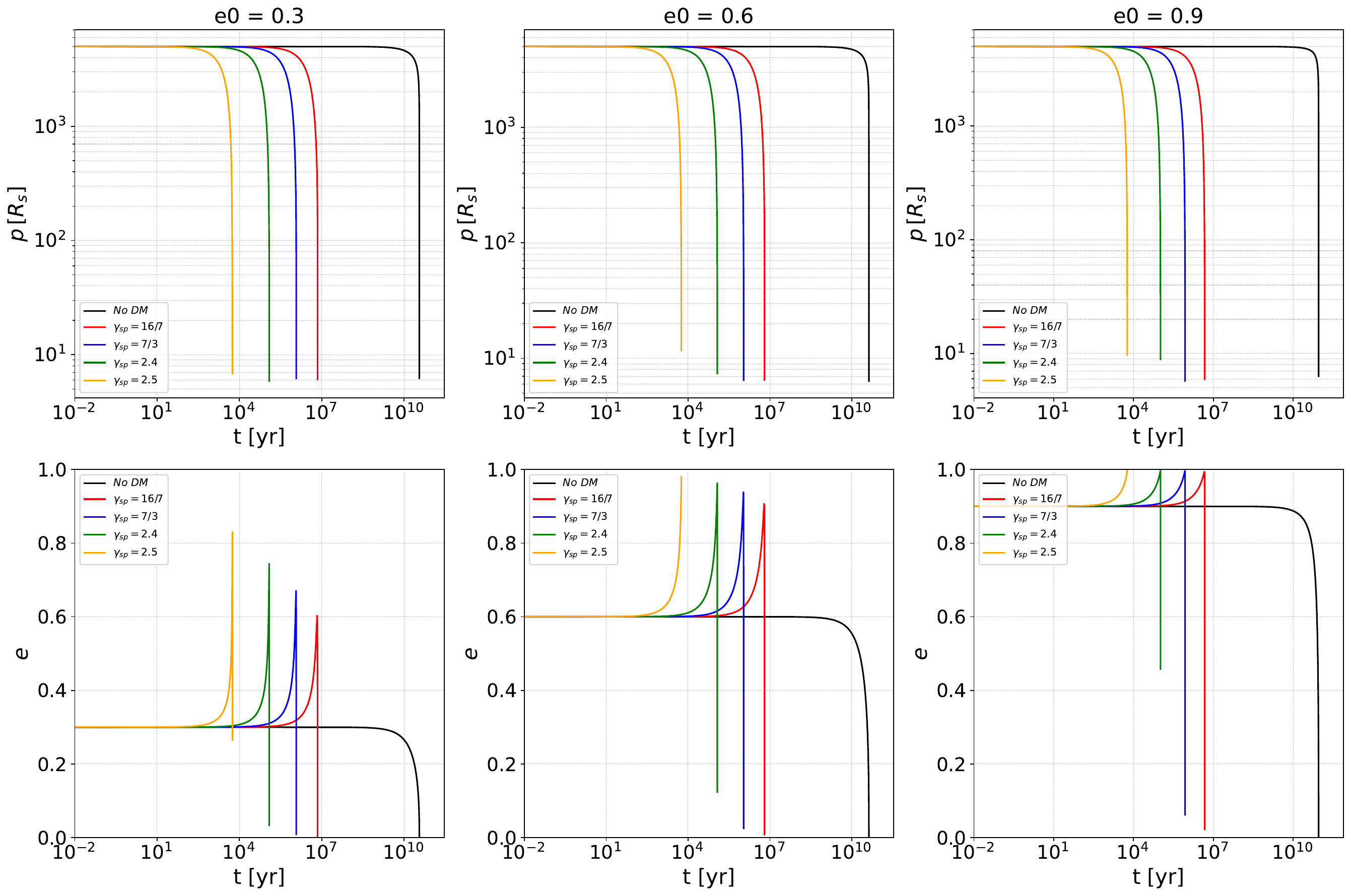}
    \caption{Orbital evolution of a secondary BH in a DM spike environment with the profile in Eq.~\eqref{p-spike_1}. 
        Initial conditions are fixed at $p_0 = 5000\,Gm_1/c^2$ with increasing initial 
        eccentricities (from left to right: $e_0 = 0.3$, $0.6$, $0.9$).}
    \label{fig:p_spike_five}

\end{figure*}

For systems with initial semi-latus rectum $p_0 = 5000\,Gm_1/c^2$, comparison of Figs.~\ref{fig:p_spike_one} and~\ref{fig:p_spike_five} shows that while increasing the initial $p$ by a factor of 5 results in a $10^3$ longer merger timescale without DM, the presence of a DM spike reduces this increase to just 1 order of magnitude, demonstrating the significant acceleration of BBH coalescence through DM dynamical friction and accretion. The environmental effects dominate over GW emission at larger separations, primarily driving eccentricity enhancement, with $\gamma_{\mathrm{sp}}=\{16/7,7/3,2.4,2.5\}$ producing eccentricity differences $\Delta e\sim\{0.3,0.4,0.5,0.6\}$ for $e_0=0.3$, while systems with $e_0=0.9$ all exhibit eccentricity growth to nearly 1 followed by rapid decay to 0. 
Thus, the DM leads to orbital decay and an increase in eccentricity. As the BBHs get closer, the GWs become dominant and cause orbital circularization. From the Figs.~\ref{fig:p_spike_one} and~\ref{fig:p_spike_five}, we can see that for $\gamma_{\mathrm{sp}}=2.5$, the orbit falls within $r_{\mathrm{ISCO}}$ before it has time to circularize. This effect is more pronounced for larger initial eccentricity $e_0$ or larger initial semi-latus rectum $p_0$.

Throughout the evolution, the DM spike formation leads to clearly distinguishable BBH dynamics for different $\gamma_{\mathrm{sp}}$ values (see Figs.~\ref{fig:p_spike_one}--\ref{fig:p_spike_five}). For gNFW profiles without spike formation around the SMBH, distinguishing DM distributions with $\gamma < 1$ would require larger initial $p_0$ parameters.

\section{Parameter Space Constraints}
\label{sec:waveform}
Our previous investigation comprehensively analyzed the full temporal evolution of both the semi-latus rectum $p$ and eccentricity $e$ in BBH systems, though the required observational timescale for such processes exceeds practical feasibility. This section instead examines the behavior of a secondary BH moving along an elliptical orbit within a DM environment, focusing on whether orbit modifications caused by DM dynamical friction and accretion over a 30-year timescale could be detectable through GW observations, while simultaneously providing constraints on the DM power-law distribution.
\subsection{The SNR in PTA observations}
The waveform of GWs is produced from the inspiral of a binary system. If such a BBH system exists at the GC with orbital periods ranging from months to years, the resulting GWs would have frequencies from $\mu$Hz to nHz, potentially detectable by PTAs. The plus and cross modes are given by~\cite{husa2009michele,article}
\begin{align}
	\label{dm-gw-x}
	h_{+}=& -\frac{2G^2 m\mu}{c^4\,p\,R} \left(1+\cos^2{\iota }\right)\bigg\{ \bigg[\cos\left(2\phi+2\omega\right)\nonumber\\
	&+\frac{5e}{4}\cos\left(\phi+2\omega\right)+\frac{e}{4}\cos\left(3\phi+2\omega\right)-\frac{e^2}{2}\cos{2\omega} \bigg]\nonumber\\
	&+\frac{e}{2}\sin^2{\iota }\left(\cos{\phi}+e\right)\bigg\},\\
	\label{dm-gw-+}
	h_{\times }=& -\frac{4G^2 m\mu}{c^4\,p\,R}\cos{\iota }\left[\sin\left(2\phi+2\omega\right)+\frac{5e}{4}\sin\left(\phi+2\omega\right)\right.\nonumber\\
	&\left.+\frac{e}{4}\sin\left(3\phi+2\omega\right)+\frac{e^2}{2}\sin{2\omega} \right],
\end{align}
where $m$ is the total mass of the system and $\mu$ denotes the reduced mass. The parameter $R$ represents the distance from the BBH's center of mass to the observer; in our analysis, this corresponds to the separation between the MW's GC and the solar system. The angle between the binary's orbital angular momentum axis and the detector's line of sight is denoted by $\iota$, while $\omega$ describes the azimuthal component of the inclination angle. For simplicity, we adopt $\iota = 0$ and $\omega = 0$ throughout this work.

\begin{figure}
    \centering
    \includegraphics[height=0.3\textheight]{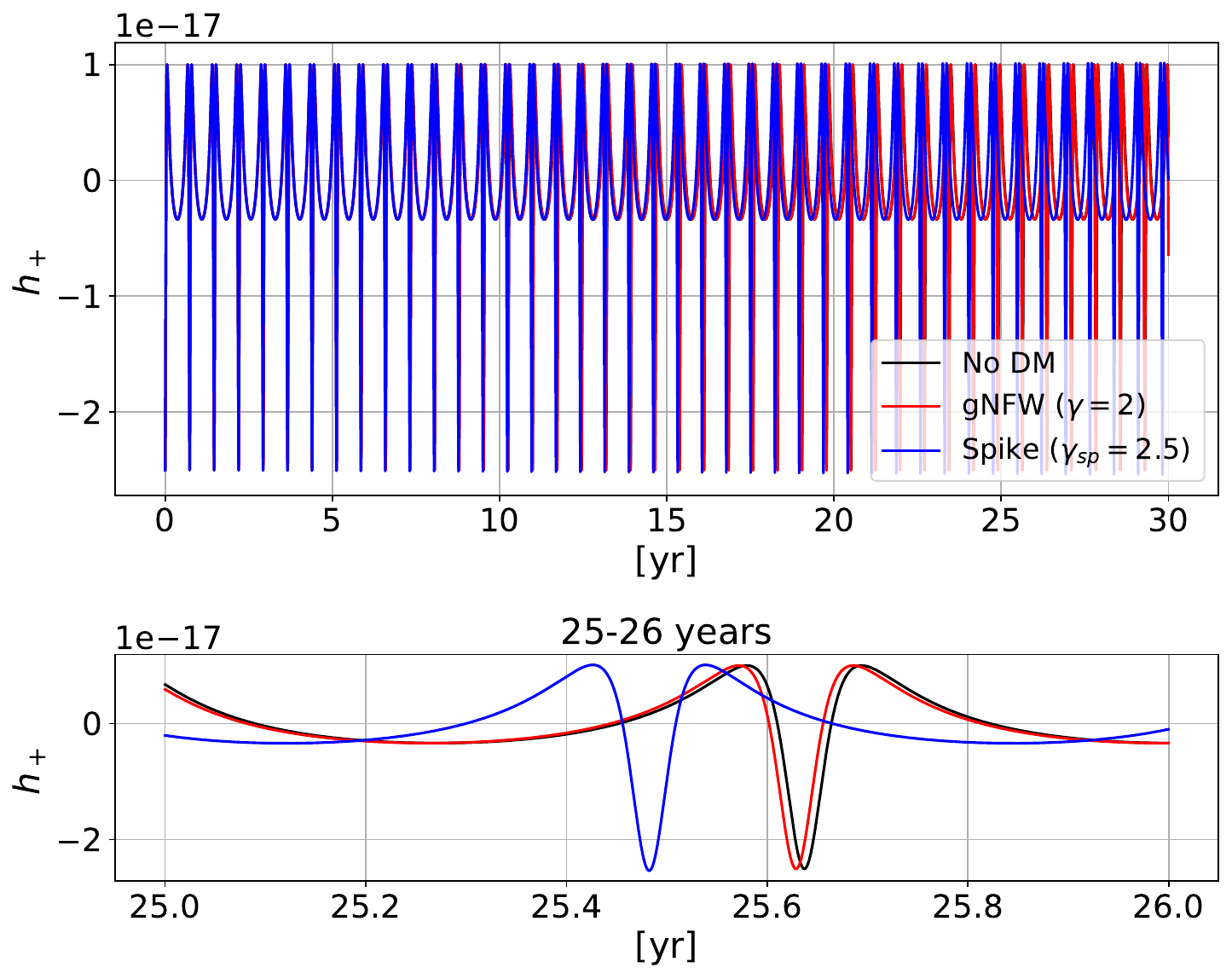}
    \caption{
        GW strain $h_+$ comparison for different DM profiles in Eq.~\eqref{eq::gnfw_halo} and Eq.~\eqref{p-spike_1}. The black line is GW waveform without the influence of DM. The red (blue) line represents GW waveform under influence of DM with gNFW (spike) density profile.
        Top: Full evolution over 30 years. 
        Bottom: Zoomed view of years 25-26.
    }
    \label{fig:strain}
\end{figure}

Fig.~\ref{fig:strain} shows the GW waveform of a binary BH system. The primary BH is the supermassive BH at the center of the MW, and the secondary BH has a mass of $1000 M_{\odot}$, with an initial eccentricity  of $e_0=0.6$ and semi-latus rectum of $p_0=2000 Gm_1/c^2$. The black solid line represents the case  without DM, the red solid line corresponds to a gNFW profile with $\gamma=2$, and the blue line shows the DM spike  distribution with $\gamma_{\mathrm{sp}}=2.5$. The top panel displays the GW over 30 years, while the bottom panel zooms in on years 25 to 26.
    
In the frequency domain, the GW waveform is given by
\begin{equation}
 \tilde{h}_{+,\times}  \left(f\right) = \int_{-\infty}^{\infty} h_{+, \times} \left(t\right) e^{2\pi ift}{{\dm}t}.
\end{equation}
Given two signals $h_1(t)$ and $h_2(t)$, we can define the inner product $\left(h_1 | h_2\right)$ as
\begin{align}
\left(h_1 | h_2\right) 
&\equiv 4 \text{Re} \int_{f_{\text{min}}}^{f_{\text{max}}}
\frac{\tilde{h}_1(f)\tilde{h}_2^{\ast}(f)}{S_n(f)} \, {{\dm}f}. \label{Eq:inner_product_def}
\end{align}
where $\text{Re}$ denotes the real part, $f_{\text{min}}=1/T_{\text{obs}}$, and $f_{\text{max}} = 1/\Delta t$. The GW frequency range that can be probed by a PTA is limited by the cadence ($\Delta t$) and the total observation period ($T_{\text{obs}}$), i.e., $1/T_{\text{obs}} \lesssim f \lesssim 1/\Delta t$. Current PTAs normally set a cadence of $\Delta t \sim 1 - 2$ weeks and have been running for a total observation period of $T_{\text{obs}} \sim 30$ years.

We define the effective signal-to-noise ratio (SNR) $\varrho$ ~\cite{Gourgoulhon:2019iyu} by
\begin{equation}\label{e:SNR_eff}
    \varrho^2 = \left(h | h\right)= 4 \int_{f_{\text{min}}}^{f_{\rm max}}\mathrm{d} f 
    \frac{|\tilde{h}_+(f)|^2+|\tilde{h}_\times(f)|^2}{S_{\rm n}(f)}  .
\end{equation}
For a PTA with $N_p$ ($\geq 3$) MSPs, two different data processing approaches exist for SNR estimation, namely the matched filtering method and the cross-correlation method. In this work, we adopt the matched filtering approach, which yields the following SNR expression~\cite{Guo:2022ilt}
\begin{equation}
\varrho^2 = \sum_{i=1}^{N_{\rm p}} 4\chi_i^2 \int_{f_{\rm min}}^{f_{\rm max}} {{\dm}f}\, \frac{|\tilde{h}_+(f)|^2 + |\tilde{h}_\times(f)|^2}{S_{{\rm n},i}(f)}.
\end{equation}
For convenience,
the total SNR used for theoretical analysis may be approximated as
\begin{equation}
\varrho^2\simeq4N_{p}\chi^2\int^{f_{\rm max}}_{f_{\rm min}}{{\dm}f}\frac{|\tilde{h}_+(f)|^2+|\tilde{h}_\times(f)|^2}{S_{\rm n}(f)}.
\label{SNR_square}
\end{equation}
By assuming that all MSPs contribute to the SNR equally~\cite{Moore:2014eua}, here, $\chi$ represents the geometric factor, which equals 0.365 under the far-field approximation~\cite{Guo:2022ilt}.

We consider the noise affecting PTA detection of individual sources to be primarily composed of two components: the shot noise and the confusion with the gravitational wave background (GWB) ~\cite{Rosado:2015epa,Goldstein:2018rdr,Chen:2020qlp}. For this analysis, we neglect the intrinsic red noises of pulsars ~\cite{Goncharov:2019mei,Lentati:2016ygu}. The power spectral density (PSD) of the GWB strain originating from shot noise is typically characterized as ~\cite{Creighton:2011zz}
\begin{equation}
	S_{\rm n,s}(f)=8\pi^2\sigma^2 f^2\Delta t,
    \label{PSD}
\end{equation}
where $\sigma$ is the root mean square (RMS) of pulsar timing residuals noise, $\Delta t$ is the mean cadence of the PTA observations and we set $N_{\rm p}=1000$, $\sigma=100$\,ns, $\Delta t=0.02$\,yr, $T=30$\,yr for SKA-PTA. The
strain of GWB due to GW radiation from numerous distant
MBBHs can be described as~\cite{Chen:2020qlp}
 
\begin{equation}
	h_{\rm b}=\mathcal{A}\frac{(f/1{\rm yr}^{-1})^{-2/3}}{[1+(f_{\rm bend}/f)^{\kappa_{\rm  gw}\gamma_{\rm gw}}]^{1/(2\gamma_{\rm gw})}}.
\end{equation}
We adopt $\log\mathcal{A}\sim-15.70$, $f_{\rm bend} =2.45\times 10^{-10}$\,Hz, $\kappa_{\rm gw}=3.74$, $\gamma_{\rm gw}=0.19$ which are the median values for the GWB predictions in ~\cite{Guo:2022ilt,Chen:2020qlp}.
The total noise
for individual PTA sources is then
\begin{equation}
	S_{\rm n}(f)=S_{\rm n,s}+\frac{h_{\rm b}^2}{f},
\end{equation}
or in the strain form as
\begin{equation}
	h_{\rm n}(f)=\sqrt{fS_{\rm n,s}+h_{\rm b}^2}.
\end{equation}
In our the SNR calculations, we treat the GWB as a noise component. 

The two waveforms are indistinguishable if the condition~\cite{fang2019impact}
\begin{equation}
\left( \delta h|\delta h \right)=\left( h_1-h_2| h_1-h_2 \right)<1 
\label{waveformdistincondition}.
\end{equation}
In practical observations, when $h_1 \approx h_2$, the aforementioned formula can be simplified. Conventionally, researchers compute a quantity called the ``Fitting Factor'' (FF)~\cite{Apostolatos:1994mx,Lindblom:2008cm}, defined as
 \begin{equation}
    \textnormal{FF}= \frac{({h}_1|{h}_2)}{\sqrt{({h}_1|{h}_1)({h}_2|{h}_2)}},
\end{equation}
and compare it with a threshold FF defined as
and compare it with a threshold \textnormal{FF} defined as

\begin{equation}
\textnormal{FFS}=1-{1\over \left(  h_1| h_1 \right)+\left(   h_2| h_2 \right)}.
\end{equation}
The quantity $\mathrm{FFS}$ is intrinsically connected to the SNR, defined as $\mathrm{SNR} = \sqrt{(h|h)}$. When $h_1 \simeq h_2$, this relationship simplifies to $\mathrm{FFS} \simeq 1 - 1/(2\,\mathrm{SNR}^2)$. Under this approximation, the criterion in Eq.~(\ref{waveformdistincondition}) reduces to $\mathrm{FF}{-}\mathrm{FFS} > 0$, which serves as the condition for waveform indistinguishability.

In BBH systems, the presence of DM significantly alters the dynamics. While pure GW radiation reaction dominates the smaller BH's motion in DM-free environments, additional effects - particularly dynamical friction and accretion - must be considered when DM is present. 

The distinguishability of gravitational waveforms depends critically on the $\mathrm{FF}{-}\mathrm{FFS}$ value:
\begin{itemize}
    \item When $\mathrm{FFS}{-}\mathrm{FF} < 0$, the GW waveforms with and without DM influence become indistinguishable.
    \item When $\mathrm{FFS}{-}\mathrm{FF} > 0$, the GW waveforms can be clearly distinguished.
\end{itemize}

\subsection{Effects of gNFW profile}

Our analysis focuses on DM's impact on the inspiral phase of the smaller BH within binary systems. Using PTA detection ranges~\cite{Guo:2024tlg}, we explore the SNR distribution across the parameter space of mass and semi-latus rectum $p$, identifying regions where DM most significantly affects the $\mathrm{FFS}-\mathrm{FF}$ relationship.
We investigate GWs from BBH using SKA-PTA observations with a total duration of 30 years. Assuming a gNFW  profile in Eq.~\eqref{eq::gnfw_halo} around the SMBH at the GC without considering spike formation, we search within the PTA sensitivity range to identify the parameter space where the effects of dynamical friction and accretion from DM are most significant while remaining detectable through GW  observations.

\begin{figure*}[t]
    \centering
    \includegraphics[width=\textwidth]{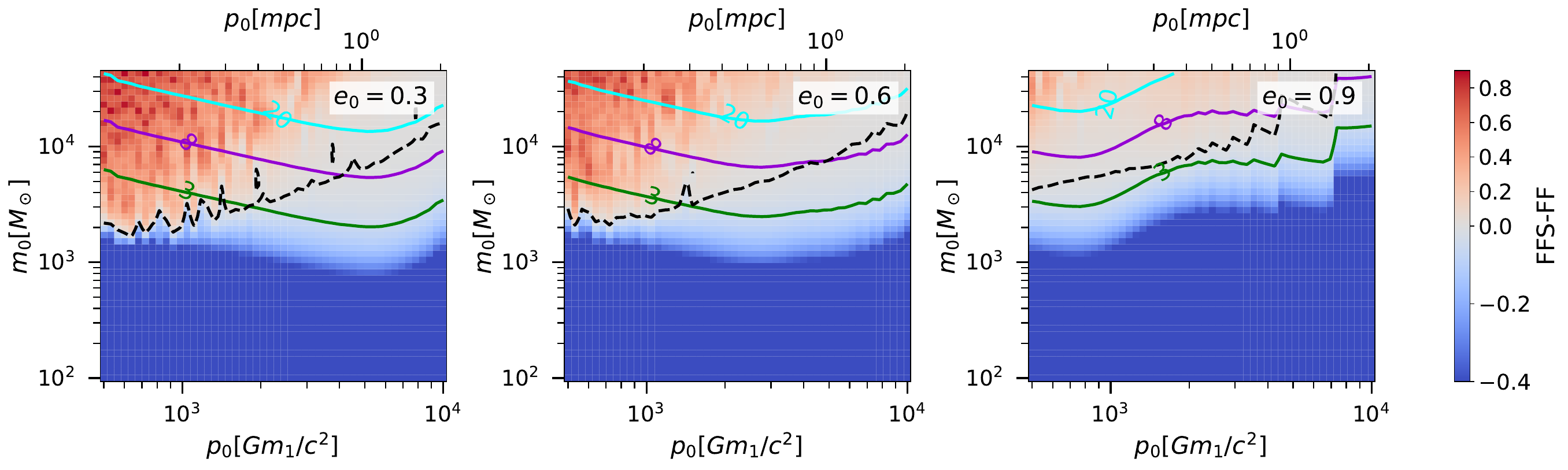}
    \caption{The secondary BHs orbiting the SMBH at the GC, where the DM distribution follows a gNFW profile in Eq.~\eqref{eq::gnfw_halo} with $\gamma=2$, emit GWs potentially detectable by SKA-PTA. The analysis shows the parameter space with the horizontal axis representing the semi-latus rectum $p_0$, in units of $R_1=Gm_1/c^2$ ranging from $500 R_1$ to $1000 R_1$ (approximately 0.1mpc to 2mpc), and the vertical axis shows the secondary BH mass $m_0$ in solar mass units, spanning from $100M_{\odot}$ to $42600\,M_{\odot}$. The top axis is the semi-latus rectum $p_0$ labeled in mpc.  Black dashed contours indicate the FFS--FF = 0 isoline. The SNR contours are displayed as green curves for SNR $= 3$, purple for SNR $= 8$, and cyan for SNR $= 15$. The three panels present results for different initial orbital eccentricities with $e_0 = 0.3$ at left, $e_0 = 0.6$ at center, and $e_0 = 0.9$ at right.
    The color gradient represents FFS--FF values, where red regions indicate FFS--FF $> 0$, blue regions show FFS--FF $< 0$, and silver-white areas correspond to FFS--FF $\approx 0$.
    }
    \label{fig:gNFW_SKA_gamma_two}
\end{figure*}

Fig.~\ref{fig:gNFW_SKA_gamma_two} displays our analysis results for the $\gamma=2$ scenario, 
illustrating the FFS-FF values via a color gradient as functions of the semi-latus rectum $p_0$ 
(on the horizontal axis) and the secondary BH mass $m_0$ (on the vertical axis). The fitting factor FF serves as a quantitative measure of waveform similarity between systems with and without DM effects, where higher FF values indicate greater similarity and consequently weaker DM influence. The threshold FFS determines the critical boundary for observable waveform differences, with red regions where FFS-FF $> 0$ representing detectable differences whose magnitude correlates with color intensity, blue regions where FFS-FF $< 0$ corresponding to negligible differences, and silvery-white areas near zero indicating borderline cases requiring extended observation periods. The black dashed contour marks the FFS-FF = 0 level while green, purple and cyan contours denote SNR of 3, 8 and 20 respectively. The three panels systematically present results for initial orbital eccentricities of 0.3 in the left panel, 0.6 in the central panel, and 0.9 in the right panel.

For Fig.~\ref{fig:gNFW_SKA_gamma_two} with initial eccentricity $e_0=0.3$, scattered black dots appear above the SNR=8 contour, corresponding to FFS-FF values approaching zero with minor fluctuations and measurement uncertainties. For the cases with eccentricities $e_0=0.3$ and $0.6$, the waveforms show significant differences when the semi-latus rectum $p_0 \lesssim 0.5\,\text{mpc}$ and mass $m_0 \gtrsim 2000\,M_{\odot}$. To detect such cases requires $m_0 \gtrsim 4000\,M_{\odot}$ for SNR $>3$, while precise observation of this event demands SNR $>8$, which in turn requires $m_0 \gtrsim 8000\,M_{\odot}$. 
Although SKA-PTA is more sensitive to slightly larger semi-latus rectum values, this region appears silver-white with FFS-FF approaching zero, making it unsuitable as our detection zone.
The most eccentric case with $e_0=0.9$ demonstrates detectable differences only for masses surpassing $10^4\,M_{\odot}$, exhibiting both diminished spatial extent and intensity of red regions along with an upward shift of the zero contour that reduces the detectable parameter space.

The analysis reveals non-monotonic color variations with noticeable fluctuations within red regions, accompanied by corresponding undulations in the zero contour. Measurement precision improves with higher SNR, and the combination of parameter space above both the zero contour and the SNR$=$8 threshold provides optimal detection conditions. Based on comprehensive examination of these results, we conclude that selecting regions with SNR exceeding 8 yields the most reliable waveform difference detection.

\begin{figure*}[t]
    \centering
    \includegraphics[width=\textwidth]{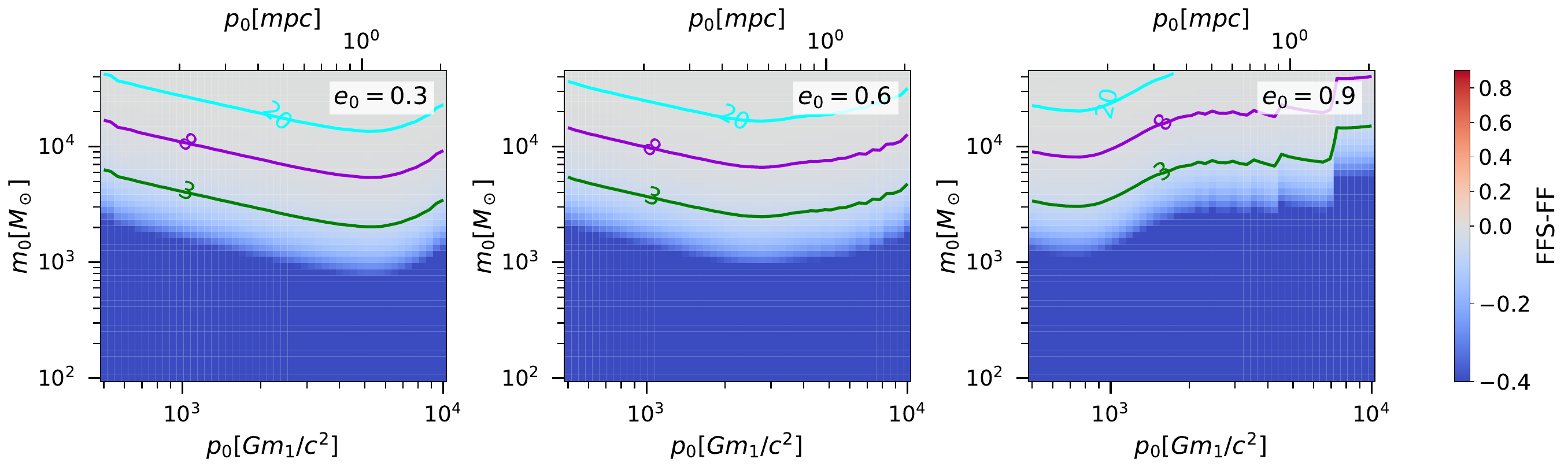}
    \caption{The detection analysis for the DM density profile with $\gamma=1.5$ indicates that across all cases of initial orbital eccentricities from $e_0=0.3$ to $0.9$, the parameter space exhibits predominantly silvery-white distribution characteristics.
The labels are the same as decribed in Fig. \ref{fig:gNFW_SKA_gamma_two}.
This observation suggests marginal detectability of waveform differences during the 30-year observation period. The figure shows FFS-FF$\lesssim0$ throughout, with no FFS-FF$=0$ contour line present.}
    \label{fig:gNFW_SKA_gamma_onefive}
\end{figure*}

For the DM distribution with $\gamma=1.5$ shown in Fig.~\ref{fig:gNFW_SKA_gamma_onefive}, we observe significant differences compared to the $\gamma=2$ case. Notably, during SKA-PTA's 30-year observation window, the entire parameter space of semi-latus rectum $p_0 \in [500\,Gm_1/c^2, 10^4\,Gm_1/c^2]$ and compact object masses $m_0 \gtrsim 1000\,M_{\odot}$ appears silvery-white. When plotting the FFS-FF=0 contour line, it does not appear in the figure because FFS-FF $\lesssim$ 0 throughout the silvery-white region, indicating no detectable waveform differences across the entire parameter space for $\gamma=1.5$.

If we could establish a GC-PTA consisting of 10 MSPs within our galaxy, it would enable exploration of a larger parameter space and increase the FFS threshold to reveal more subtle waveform differences. The detection results using GC-PTA are presented in the Appendix \ref{sec:GCPTA}. 
As shown in the gNFW profile with $\gamma=2$ in Fig.\ref{fig:gNFW_GC_gamma_two} in the Appendix \ref{sec:GCPTA}, using GC-PTA detection, a mass close to $100M_{\odot}$ can be detected at a semi-latus rectum around 0.2 mpc with an eccentricity of 0.3, while a mass of approximately $200M_{\odot}$ can be detected at an eccentricity of 0.6. For an eccentricity of 0.9, a mass of $5000M_{\odot}$ can be detected. GC-PTA significantly expands the detectable parameter space.
For the distribution with $\gamma=1.5$, we cannot detect the waveform differences even when using GC-PTA, so we do not present it in the Appendix \ref{sec:GCPTA}.

\subsection{Effects of DM Spike Profile}

\begin{figure*}[t]
    \centering
    \includegraphics[width=\textwidth]{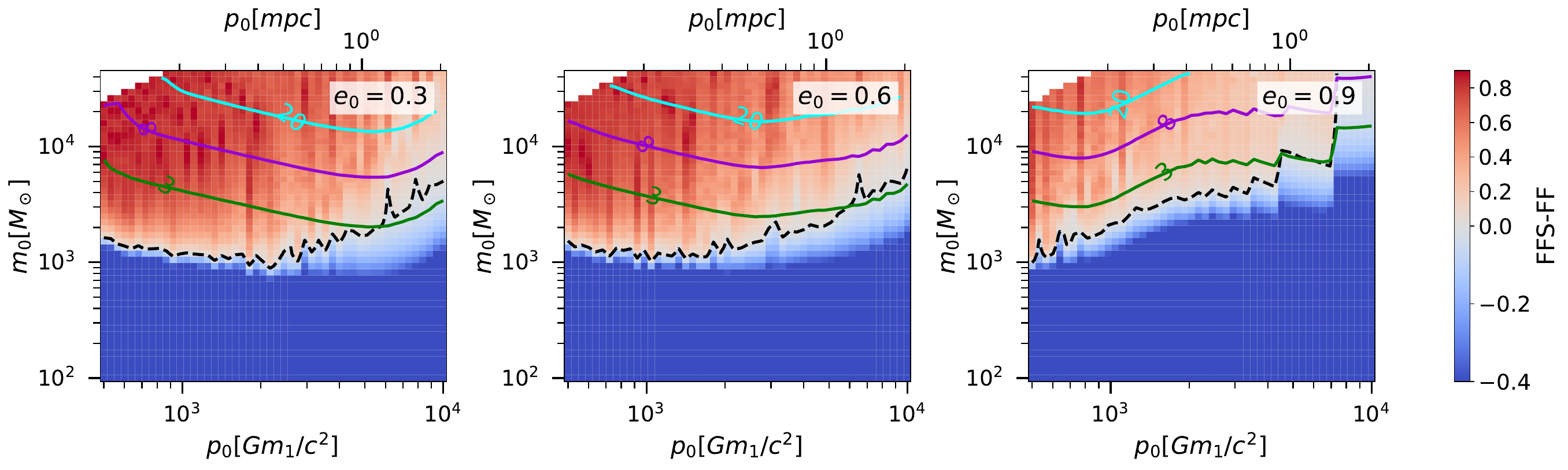}
\caption{We consider a small BH in an elliptical orbit within a DM spike characterized by a power-law slope of  $\gamma_{\mathrm{sp}}=2.5$ in Eq.~\eqref{p-spike_1}, which formed by a gNFW profile with $\gamma=2$ in Eq.~\eqref{eq::gnfw_halo}. 
The horizontal axis shows the semi-latus rectum $p_0$, while the vertical axis represents the mass $m_0$ of the small BH. 
The green, purple, and cyan contour lines correspond to SNR values of 3, 8, and 20, respectively. The white translucent region in the upper left corner indicates where the small BH's orbit radius $r=p/(1+e\cos\phi)<r_{\mathrm{ISCO}}$ within the 30-year evolution timescale.
The color bar represents the FFS-FF, where red regions indicate significant detectable differences in gravitational waveforms, blue regions correspond to undetectably small differences, and silver-white regions represent the detection threshold, the dashed line represents the FFS-FF$=0$ contour.
    }
    \label{fig:spike_ska_color_two}
\end{figure*}

This study investigates DM spike structures in Fig.~\ref{fig:spike} formed through adiabatic compression around the MW's central BH, a process that significantly enhances the DM density distribution. For spikes $\gamma_{\mathrm{sp}}=2.5$ generated from gNFW profiles with $\gamma=2$, Fig.~\ref{fig:spike_ska_color_two} shows that compared to the no-spike scenario, the red region expands and darkens, indicating more pronounced waveform differences in GWs from BBH systems under DMr's gravitational influence.
The transparent area in the figure's upper left corner corresponds to binary systems reaching $r < r_{\mathrm{ISCO}}$ within 30 years, which we exclude from analysis. As the BBHs inspiral over time, relativistic effects in dynamical friction and accretion become significant when their separation falls below \(30 \, Gm_1/c^2\) \cite{2025arXiv250509715V}. However, due to the observational constraints of PTA,  the maximum detectable 
GW frequency is \(f_{\text{max}} = 1/\Delta t\) with \(\Delta t \simeq 0.02 \, \text{yr}\). Thus, the smallest measurable orbital separation is approximately \(r_{\text{min}} \simeq 450 \, Gm_1/c^2\), which is much larger than \(30 \, Gm_1/c^2\). This ensures that the data remain largely unaffected by inaccuracies from strong-field relativistic effects, thereby validating the use of the weak-field approximation.

For the initial eccentricity \( e_0 = 0.3 \), SKA-PTA is most sensitive at 1 mpc. Around this region, we observe that a mass \( m_0 \gtrsim 2000M_{\odot} \) yields SNR $>$ 3, while a precise detection with SNR = 8 requires \( m_0 \gtrsim 5000M_{\odot} \).  
For the initial eccentricity \( e_0 = 0.6 \), the highest sensitivity occurs at 0.5 mpc, where a mass \( m_0 \gtrsim 2000M_{\odot} \) is detectable, and a precise detection requires \( m_0 \gtrsim 6000M_{\odot} \).  
For the initial eccentricity \( e_0 = 0.9 \), the optimal sensitivity is at 0.2 mpc, where a mass \( m_0 \gtrsim 3000M_{\odot} \) achieves SNR = 3, while a mass \( m_0 \gtrsim 7000M_{\odot} \) is needed for SNR = 8.

\begin{figure*}[t]
    \centering
        \includegraphics[width=\textwidth]{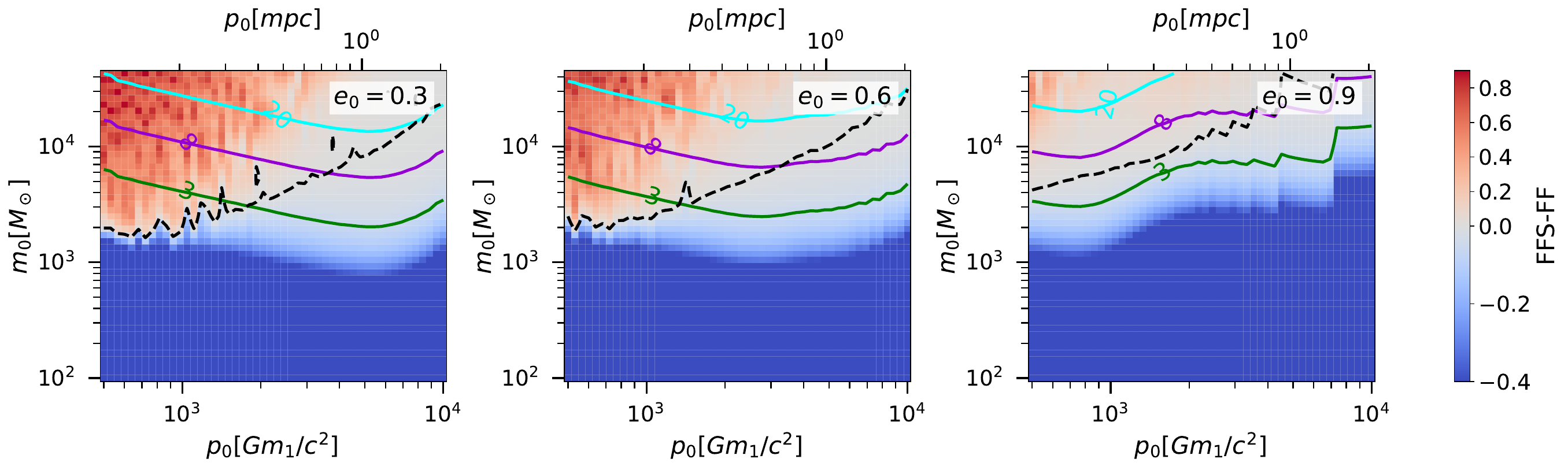}
        \caption{DM density profile with $\gamma=1.5$ under the gNFW profile in Eq.~\eqref{eq::gnfw_halo}, forming a spike with $\gamma_{\rm sp}=2.4$ in Eq.~\eqref{p-spike_1}, due to adiabatic compression by the central SMBH. 
        The horizontal axis represents the semi-latus rectum $p$ of the orbit, while the vertical axis shows the mass of the small BH. 
        The color bar indicates the FFS-FF (GW waveform mismatch), with green, purple, and cyan contour lines corresponding to SNR of 3, 8, and 20, respectively.}
        \label{fig:spike_ska_color_one_five}
\end{figure*}

The case of a spike with $\gamma_{\mathrm{sp}}=2.4$ is shown in Fig.~\ref{fig:spike_ska_color_one_five}. For $e_0=0.3$, the area above the dashed line represents waveform differences, where scattered black dots in the white region appear due to FF approaching unity and FFS also nearing unity at sufficient SNR, causing FFS-FF values to fluctuate around zero. 
Compared to $\gamma_{\mathrm{sp}}=2.5$, the $\gamma_{\mathrm{sp}}=2.4$ case shows that for eccentricities $e_0=0.3$ and $0.6$, the black dashed line imposes constraints, making it impossible to detect waveform differences at SKA-PTA's most sensitive regions. The area of the red zone is significantly reduced, and its depth is also diminished. A mass $m_0 \gtrsim 4000M_{\odot}$ near a semi-latus rectum of 0.2 mpc can achieve SNR $\gtrsim 3$, while a mass $m_0 \gtrsim 7000M_{\odot}$ around 0.5 mpc reaches SNR$ \gtrsim 8$. For $e_0=0.9$, since the black dashed line lies above the SNR=3 contour, the required mass is $m_0 \gtrsim 10^4M_{\odot}$ at $p_0 \lesssim 0.2$ mpc.

\begin{figure*}[t]
    \centering
        \includegraphics[width=\textwidth]{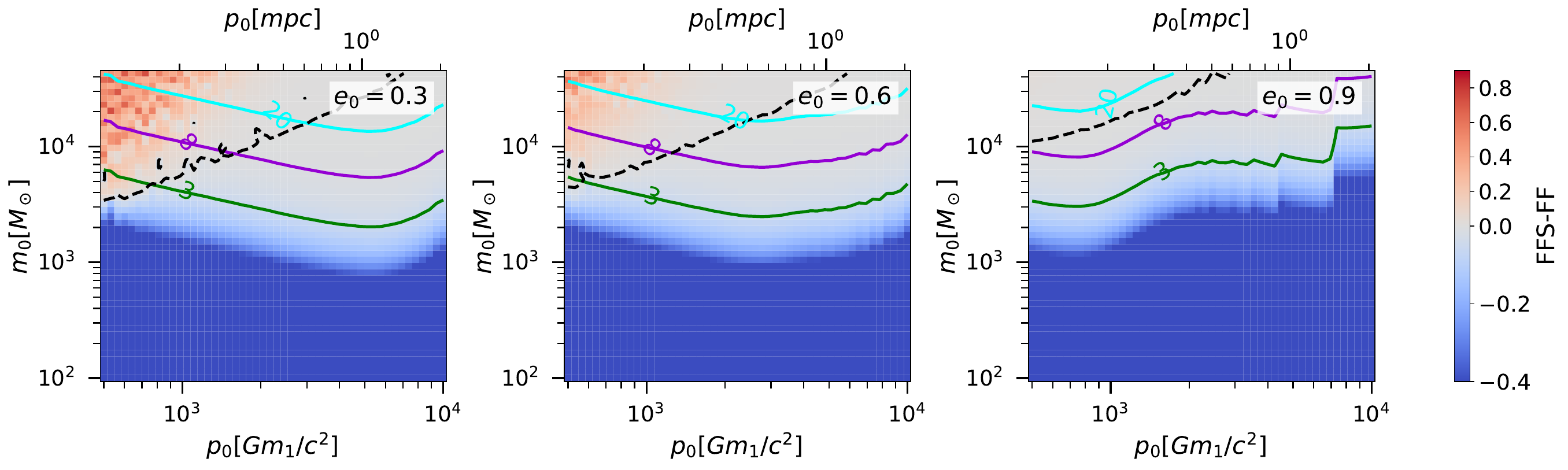}
        \caption{DM density profile with $\gamma=1$ under the gNFW profile in Eq.~\eqref{eq::gnfw_halo}, forming a spike with $\gamma_{\rm sp}=7/3$ due to adiabatic compression by the central SMBH. 
        The horizontal axis represents the semi-latus rectum $p$ of the orbit, while the vertical axis shows the mass $m_0$ of the small BH. 
        The color bar indicates the FFS-FF (GW waveform mismatch), with green, purple, and cyan contour lines corresponding to SNR of 3, 8, and 20, respectively.}
        \label{fig:spike_ska_color_one}
\end{figure*}

We initially omitted the results for $\gamma=1$ and $\gamma=0$ cases because their lower DM densities resulted in FFS-FF\,$<0$ throughout our parameter space, rendering the DM effects undetectable within a 30-year observation window. Fig.~\ref{fig:spike_ska_color_one} displays the post-spike formation scenario for $\gamma=1$ ($\gamma_{\rm sp}=7/3$). For systems with eccentricities of 0.3 and 0.6, waveform differences become observable when $m_0\gtrsim4000M_{\odot}$ near semi-latus rectum values of 0.2mpc. Reliable detection within SNR$\ge 8$ requires satisfying both the constraint $p_0\lesssim 0.4\rm mpc$ imposed by the dashed line and maintaining masses above $10^4M_{\odot}$. For \( e_0 = 0.9 \), we observe that the dashed line has shifted upward above the SNR=8 contour. Moreover, the region above the dashed line appears entirely silver-white, showing no discernible waveform differences in the plot.

\begin{figure*}[t]
        \includegraphics[width=\textwidth]{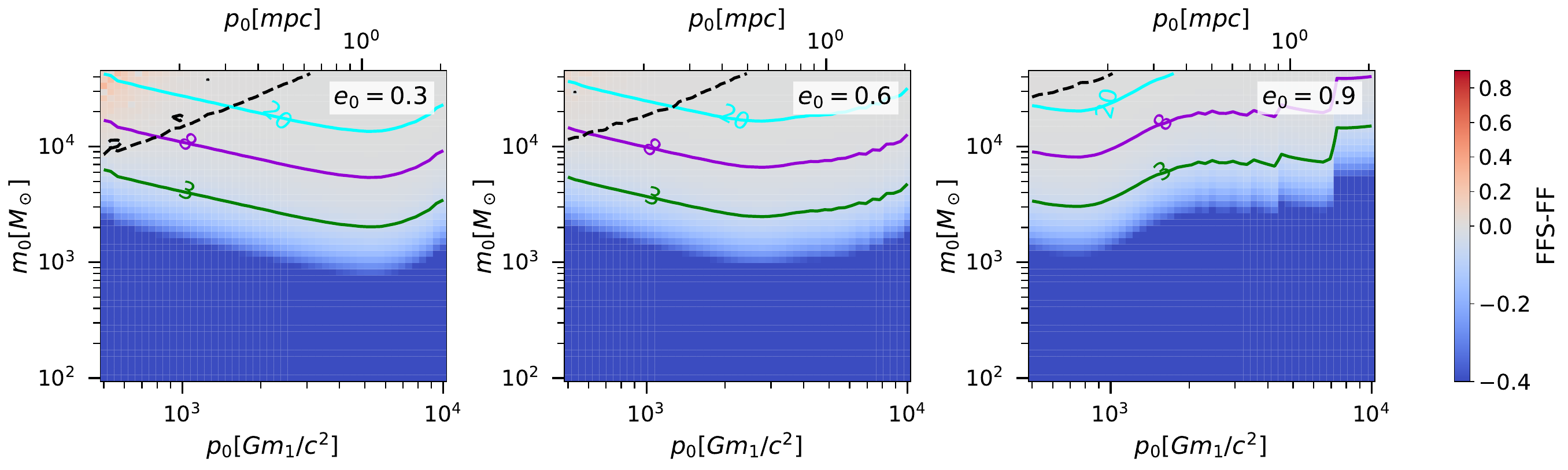}
    \caption{DM density  profile with $\gamma=0.5$ under the gNFW profile in Eq.~\eqref{eq::gnfw_halo}, forming a DM spike  in Eq.~\eqref{p-spike_1} with $\gamma_{\rm sp}=16/7$ due to adiabatic compression by the central SMBH. 
The horizontal axis represents the semi-latus rectum $p_0$ of the orbit, while the vertical axis shows the mass $m_0$ of the small BH. 
The color bar indicates the FFS-FF (GW waveform mismatch), with green, purple, and cyan contour lines corresponding to SNR of 3, 8, and 20, respectively. }
    \label{fig:spike_ska_color_zero_five}
\end{figure*}

Fig.~\ref{fig:spike_ska_color_zero_five} presents the post-spike scenario for $\gamma=0.5$ with a resultant power-law index $\gamma_{\rm sp}=16/7$. In this case, detectable waveform differences emerge only in the upper-left parameter space region where $m_0 > 2\times10^4\,M_{\odot}$ and $p_0 \lesssim 0.2\rm mpc$ for the initial eccentricity $e_0=0.3$. 
For higher eccentricities, the entire detectable parameter space within SKA-PTA's sensitivity appears silvery-white, indicating that longer observation times would be required. Our results demonstrate that if the MW's initial DM distribution followed $\gamma=0.5$, the combined effects of dynamical friction and accretion from the resulting spike profile would remain undetectable within a 30-year observation window.

In the Appendix \ref{sec:GCPTA}, we present GC-PTA detection results for spike profiles with $\gamma_{\mathrm{sp}}=2.5$ and $2.4$, as shown in Figs.~\ref{fig:spike_GC_color_two} and~\ref{fig:spike_GC_color_one_five} respectively. For eccentricities $e_0=0.3$ and $0.6$, we can detect BHs with masses $m_0 \sim 200M_{\odot}$ at orbital semi-latera recta around $p_0=0.2$ mpc.
For the cases with $\gamma_{\mathrm{sp}}=7/3$ and $16/7$, as shown in Figs.~\ref{fig:spike_GC_color_one} and~\ref{fig:spike_GC_color_zero_five}, the dashed constraints shift downward compared to SKA-PTA results, expanding the parameter space where FFS-FF $>0$. However, due to weaker DM effects (FF $\approx$ 1), the broadened parameter space remains silver-white in the plots, indicating no observable waveform differences.

\section{Conclusions}
\label{sec:concl}

In this work, we investigate the influence of DM environments on the elliptical orbits of BBHs to probe the DM distribution at the GC. Orbital modifications due to DM induce deviations in the gravitational waveform, with our analysis focusing primarily on the PTA frequency band.
We examine various power-law inner slopes of the gNFW profile, as well as the resulting DM distribution after spike formation, and identify which slopes allow detectable DM effects within a 30-year observation window using SKA-PTA. Constraints are derived for these slope parameters. For detectable cases, we further constrain the parameter space of low-mass BHs.

In particular, we investigate the density profile of DM with different power-law indices $\gamma$ in gNFW profile in Eq.~\eqref{eq::gnfw_halo}. 
Under these constraints that, the total DM mass within the MW's virial radius $r_{200}$ is $M_{\text{DM}}^{200} \approx 1.0 \times 10^{12} \, M_{\odot}$, and a local DM density near the Solar System is $\rho(r_{\odot}) = \rho_{\odot} = 0.4 \, \text{GeV/cm}^3$, we consider the adiabatic compression of DM within the gravitational influence radius of the SMBH at the GC, which forms a DM spike in Eq.~\eqref{p-spike_1} and significantly enhances the DM distribution in this region. 
We subsequently delve into the scenario of elliptical orbits, in which a secondary BH situated within a DM environment undergoes alterations in both its orbital semi-latus rectum $p$ and eccentricity $e$ as a result of GW radiation.
Simultaneously, the orbit is also affected by dynamical friction and DM accretion, which further alter these orbital parameters. 

Our analysis shows that while GW radiation tends to decrease orbital eccentricity $e$, the presence of DM increases it. For the same initial eccentricity $e_0$, this enhancement becomes more pronounced with higher DM densities and larger initial semi-latus rectum values. For initial semi-latus rectum values of $p_0 = 1000 \, Gm_1/c^2$$(\sim0.2\rm mpc)$ and $5000 \, Gm_1/c^2$$(\sim1\rm mpc)$, we observe that the orbital evolution under gNFW profiles in Eq.~\eqref{eq::gnfw_halo} with $\gamma = 1$ and $0.5$ shows little difference compared to the DM-free case. However, when a DM spike in Eq.~\eqref{p-spike_1} is present, significant differences in the orbital evolution become apparent.

The DM environment exerts dynamical friction and accretion effects on orbiting BHs, causing deviations from DM-free orbital motion that manifest in gravitational waveforms. We examine whether different DM distributions affect secondary BHs detectably within 30-year observations, identifying parameter spaces (secondary BH masses and semi-latus rectum values) where waveform differences are most observable.
To quantify detectability, we compute the frequency-domain inner product  between waveforms with and without DM effects. By setting a threshold FFS, we determine when differences become observable, and present SNR contours delineating detectable parameter spaces. Our primary analysis uses SKA-PTA, with GC-PTA results in Appendix \ref{sec:GCPTA}.
Our study using SKA-PTA over a 30-year observation period demonstrates that detecting DM effects requires both FFS-FF $> 0$ and SNR $\gtrsim 8$, while favoring smaller masses for the secondary BH probe. The detection thresholds for different DM distributions are as follows:

For the gNFW profile in Eq.~\eqref{eq::gnfw_halo} with $\gamma=2$, systems with eccentricities $e_0=0.3$ and $0.6$ require $m_0 \gtrsim 8000 M_{\odot}$ when $p_0 \approx 0.5$ mpc.  While for $e_0=0.9$, the same mass threshold applies but only when $p_0 \lesssim 0.2$ mpc. Other gNFW profiles remain undetectable within the 30-year timeframe across the entire parameter space.
For spike profiles in Eq.~\eqref{p-spike_1}, the results show a strong dependence on $\gamma_{\mathrm{sp}}$:
When $\gamma_{\mathrm{sp}}=2.5$, the minimum mass requirements are $m_0 \gtrsim 5000 M_{\odot}$ for $e_0=0.3$ ($p_0 \approx 1$ mpc), $m_0 \gtrsim 6000 M_{\odot}$ for $e_0=0.6$ ($p_0 \approx 0.5$ mpc), and $m_0 \gtrsim 7000 M_{\odot}$ for $e_0=0.9$ ($p_0 \approx 0.2$ mpc). 
For $\gamma_{\mathrm{sp}}=2.4$, these increase to $m_0 \gtrsim 7000 M_{\odot}$ ($e_0=0.3$, $p_0 \approx 0.5$ mpc), $m_0 \gtrsim 7000 M_{\odot}$ ($e_0=0.6$, $p_0 \approx 0.4$ mpc), and $m_0 \gtrsim 8000 M_{\odot}$ ($e_0=0.9$, $p_0 \approx 0.2$ mpc).
The steeper $\gamma_{\mathrm{sp}}=7/3$ profile demands $m_0 \gtrsim 10^4 M_{\odot}$ for $e_0=0.3$ and $0.6$ when $p_0 \approx 0.2$ mpc, while showing no detectable signal for $e_0=0.9$.
For the extreme case of $\gamma_{\mathrm{sp}}=16/7$, detection is only possible for $e_0=0.3$ with $p_0 \approx 0.1$\,mpc, requiring $m_0 \gtrsim 2\times10^4 M_{\odot}$, with other eccentricities remaining undetectable.

These findings highlight how detection feasibility depends critically on both the DM density profile slope and the orbital parameters, with steeper profiles and higher eccentricities generally requiring more massive perturbers in tighter orbits for successful detection.
For GC-PTA, when $\gamma=2$ in the gNFW profile with eccentricities $e_0=0.3, 0.6$, the mass can be reduced to around $300M_{\odot}$ near $p_0=0.2$ mpc. For spikes with $\gamma_{\mathrm{sp}}=2.5, 2.4$, GC-PTA can detect even smaller BHs. For other distributions, GC-PTA does not show significant advantages.

During the 30-year inspiral evolution of the BBH system, the continuous accretion of DM by the secondary BH leads to a gradual reduction in the DM density. To evaluate the impact of accretion on the DM density, we referred to the results from \cite{Karydas:2024fcn} and found that within the orbital range detectable by PTA in our current study, the resulting change in density due to accretion is negligible. For example, we  estimated the mass of DM accreted by a secondary BH of mass \(m_2=10^4 \, M_{\odot}\) inspiraling over 30 years from an initial orbital semi-latus rectum  of \(p_0=1000 \, Gm_1/c^2\) to its final location in a spike profile with $\gamma_{\rm sp} = 2.5$. The calculation indicates that the corresponding average change in DM density between these two orbits is approximately $10^{-6}$, which also demonstrates that the influence of DM accretion on the density distribution is negligible.

When the binary separation becomes sufficiently small, relativistic effects in accretion and dynamical friction become significant. However, these regimes are not detectable by PTA observations, and thus the corresponding data are not included in this study. Nevertheless, we consider this physical process to be of great importance. In future work, we plan to incorporate relativistic dynamical friction and accretion effects, along with the feedback from accretion on the DM density distribution. The cumulative impact of these effects over time may become non-negligible, and we intend to further investigate the DM environment around binary BHs, which is both valuable and intriguing.

Despite these simplifications, our study provides explicit criteria for identifying detectable DM imprints, thereby enabling observational constraints on the underlying DM distribution. Although a secondary massive BH may not exist in the Galactic Center, such systems could be present in the centers of other nearby galaxies, including M31, M32, M33, the Large Magellanic Cloud, M87, and others~\cite{Guo:2022ilt,Guo:2024tlg,Chen:2020qlp, 2025arXiv250602937F}. The same methods can be  readily extended to the center of other nearby galaxies to constrain the inner slope of the DM profile.

The GW signals from massive BBHs in nearby GCs offer an independent probe of the inner DM density profile slope. These observations can simultaneously constrain DM models and shed light on the nature of DM. Future studies could combine our results with galactic rotation curves and gravitational lensing data to provide multi-messenger constraints on DM properties.

\acknowledgments 
This work is supported by the National Key Research and Development Program of China (Nos. 2021YFC2203001, 2021YFC2201901), the National Natural Science Foundation of China (No.12375059),  the fellowship of China National Postdoctoral Program for Innovative Talents (Grant No. BX20230104), and the Project of National Astronomical Observatories, Chinese Academy of Sciences (No. E4TG6601).


\appendix
\section{Galactic Center Pulsar Timing Array (GC-PTA)}
\label{sec:GCPTA}
\begin{figure*}[t]
    \centering
    \includegraphics[width=\textwidth]{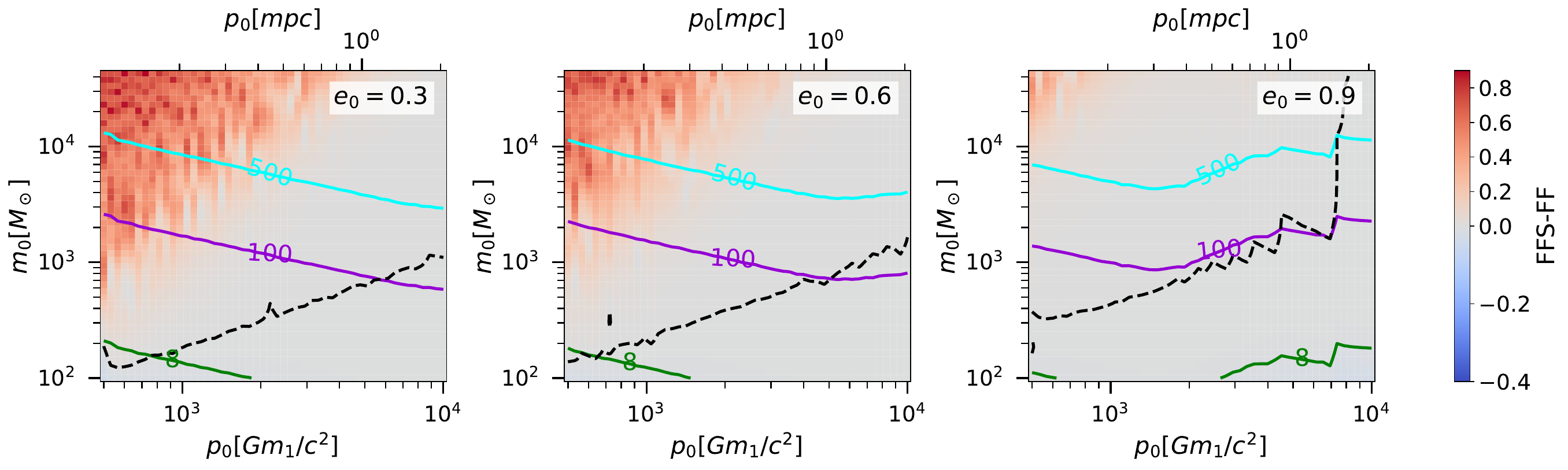}
\caption{The parameter space for secondary BH detection using GC-PTA under the gNFW DM profile in Eq.~\eqref{eq::gnfw_halo} with power-law slope $\gamma=2$. 
The x-axis shows the semi-latus rectum $p_0$ while the y-axis represents mass $m_0$. 
The FFS-FF metric quantifies GW waveform distortion significance caused by DM perturbations to BH orbits.
Color coding indicates detectability over a 30-year observation period: 
red (detectable), silver (marginal detection threshold), and blue (undetectable).     }
    \label{fig:gNFW_GC_gamma_two}
\end{figure*}

If MSPs exist within the MW, we can discover them using SKA  and utilize them as PTA  to detect low-frequency GWs.
 We designate this configuration as GC-PTA , which offers significantly higher SNR compared to conventional extragalactic PTAs. We assume there are 10 MSPs in the MW. The SNR is calculated using Eq.~\eqref{SNR_square} with $N_p=10$, where the geometric factor $\chi$ is given by
$\chi \approx 0.365\left(\frac{r}{r_{pl}}\right) \approx 2920\left(\frac{r}{8\,\text{kpc}}\right)\left(\frac{r_{pl}}{1\,\text{pc}}\right)^{-1}$~\cite{Guo:2022ilt}.
Here we assume that all 10 MSPs are located at distances $r \sim 1\,\text{pc}$ from the GC. In Eq.~\eqref{PSD}, we adopt the parameters $\sigma=100{\rm ns}$, $\Delta t=0.02{\rm yr}$, and $T=30{\rm yr}$. 
In this section, the figures demonstrate the waveform variations FFS-FF and SNR for GWs from the secondary BHs within different DM density profiles, as detected by the GC-PTA.

As shown in Fig.~\ref{fig:gNFW_GC_gamma_two},
if we do not consider DM spike formation, the GC-PTA demonstrates superior detection capabilities with higher SNR. Compared to SKA-PTA, GC-PTA can probe a larger parameter space within the 30-year observation window, with no undetectable regions (blue) in our selected parameter space. 
The analysis reveals significant DM effects for $\gamma=2$, while for $\gamma=1.5$ the FFS-FF values remain near the detection threshold indicating weaker DM influence. For even smaller power-law indices, the effects become negligible and are therefore not presented in Appendix \ref{sec:GCPTA}.

\begin{figure*}[t] 
        \includegraphics[width=\textwidth]{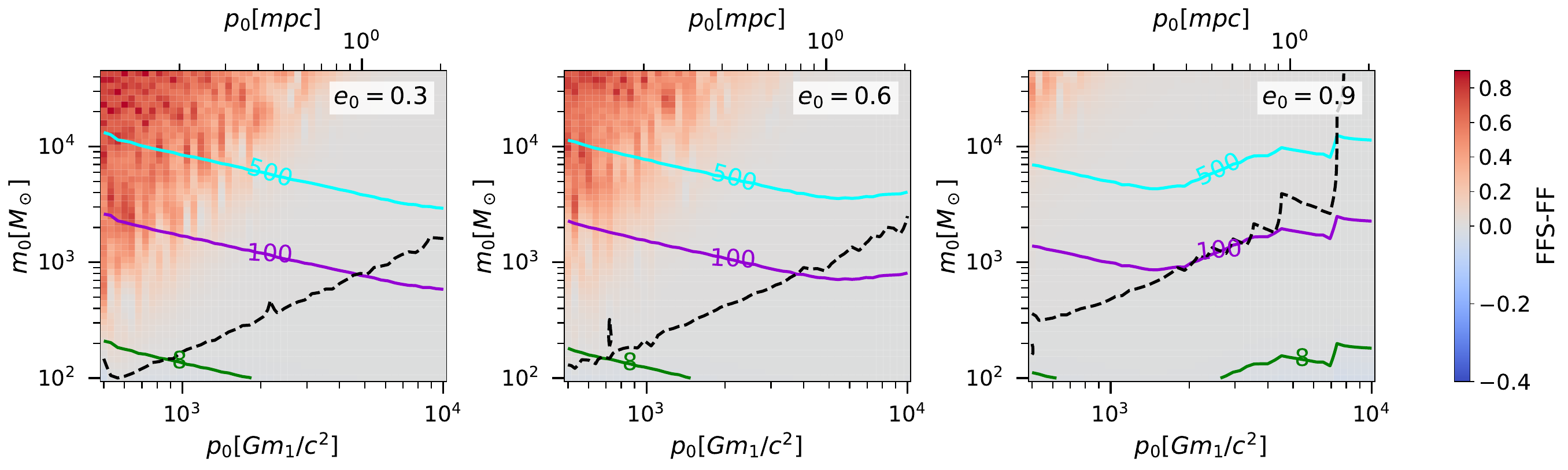}
        \caption{FFS-FF and SNR of GWs by the GC-PTA, for the DM spike  in Eq.~\eqref{p-spike_1} with $\gamma_{\mathrm{sp}}=2.4$, 
        formed by the gNFW profile with $\gamma=1.5$ in Eq.~\eqref{eq::gnfw_halo}.}
        \label{fig:spike_GC_color_one_five}
\end{figure*}

\begin{figure*}[t] 
        \includegraphics[width=\textwidth]{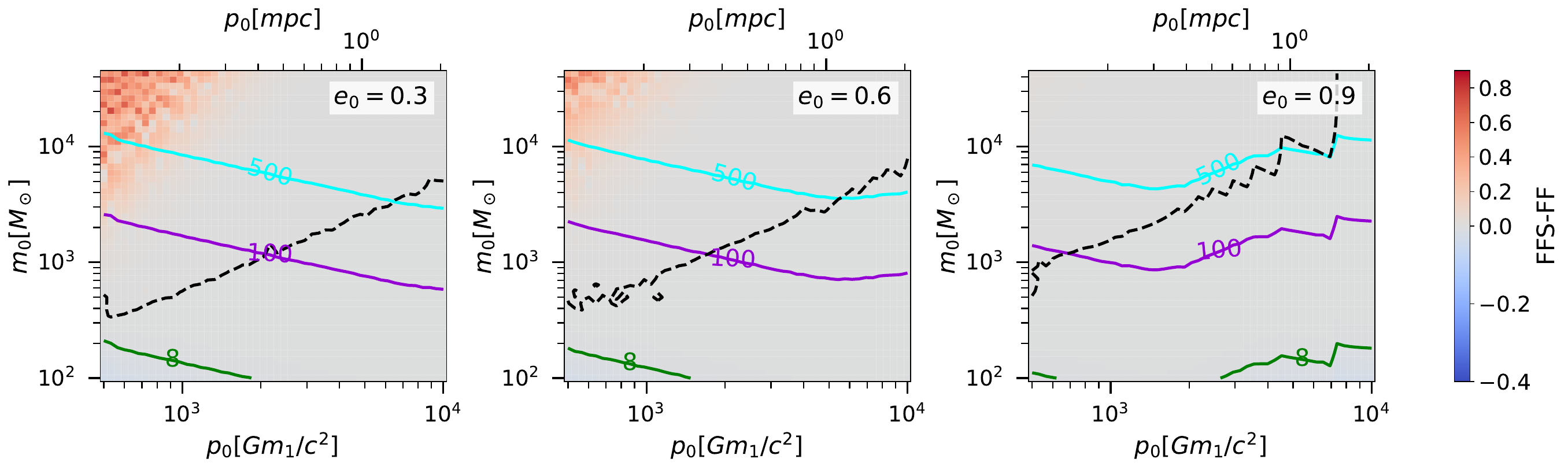}
        \caption{FFS-FF and SNR of GWs by the GC-PTA,  for the DM spike in Eq.~\eqref{p-spike_1} with $\gamma_{\mathrm{sp}}=7/3$, 
        formed by the gNFW profile with $\gamma=1$ in Eq.~\eqref{eq::gnfw_halo}.
}
        \label{fig:spike_GC_color_one}
\end{figure*}

\begin{figure*}[t] 
        \includegraphics[width=\textwidth]{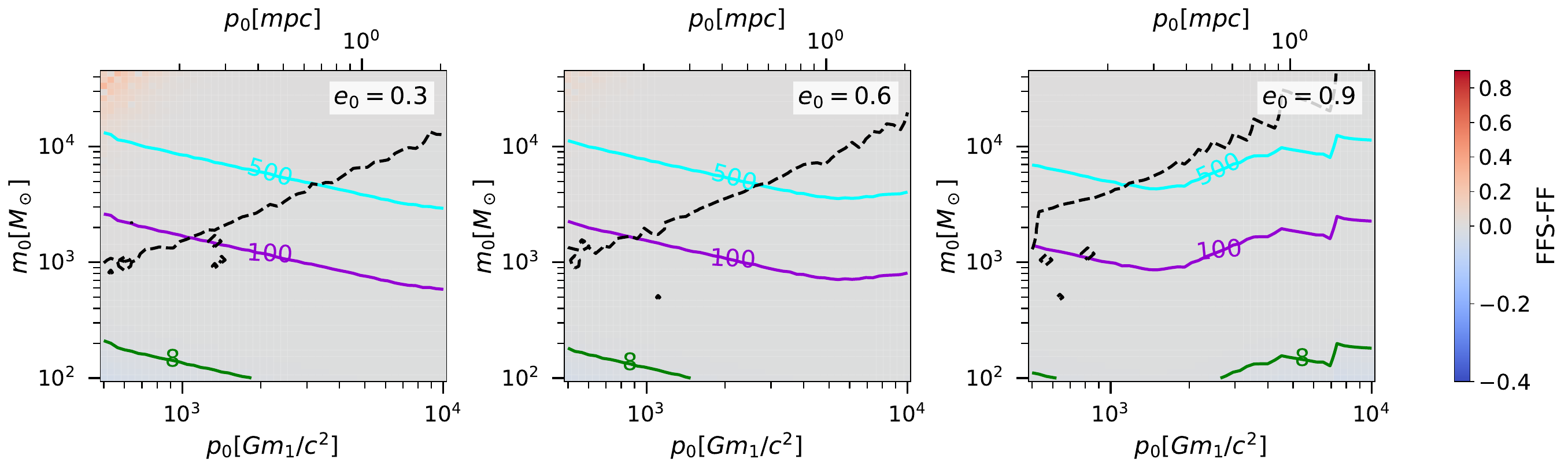}
        \caption{FFS-FF and SNR of GWs by the GC-PTA,  for the DM spike in Eq.~\eqref{p-spike_1} with $\gamma_{\mathrm{sp}}=16/7$, 
        formed by the gNFW profile with $\gamma=0.5$ in Eq.~\eqref{eq::gnfw_halo}.}
        \label{fig:spike_GC_color_zero_five}
\end{figure*}

As shown in Figs.~\ref{fig:spike_GC_color_two}, \ref{fig:spike_GC_color_one_five}, \ref{fig:spike_GC_color_one}, and~\ref{fig:spike_GC_color_zero_five}, these represent spikes formed from gNFW profiles with $\gamma=\{2,1.5,1,0.5\}$ respectively, resulting in power-law indices $\gamma_{\mathrm{sp}}=\{2.5,2.4,7/3,16/7\}$.
When DM forms a spike in Eq.~\eqref{p-spike_1}, its density and power-law index increase significantly. Compared to SKA-PTA, the GC-PTA demonstrates the capability to detect the entire selected parameter space with significantly broader coverage.

\begin{figure*}[t] 
    \centering
    \includegraphics[width=\textwidth]{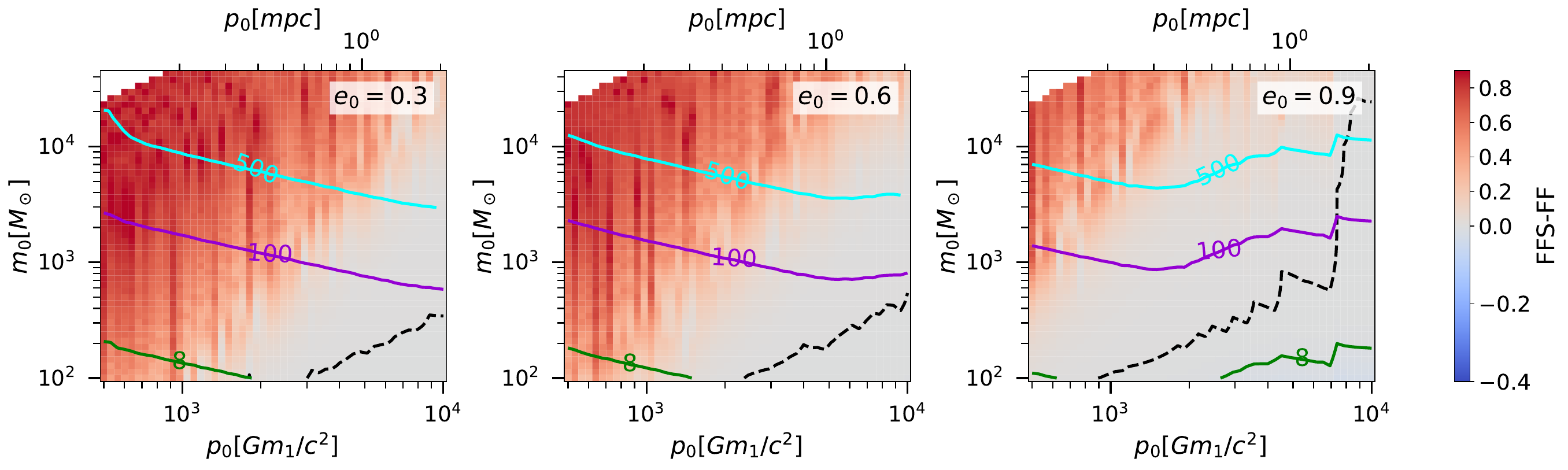}
    \caption{FFS-FF and SNR of GWs by the GC-PTA, for the DM spike in Eq.~\eqref{p-spike_1} with $\gamma_{\mathrm{sp}}=2.5$, formed by the gNFW profile with $\gamma=2$ in Eq.~\eqref{eq::gnfw_halo}.}
    \label{fig:spike_GC_color_two}
\end{figure*}
    
\bibliography{refer.bib}

\end{document}